\documentclass[journal=jacsat,manuscript=article]{achemso}

\usepackage[english]{babel}
\usepackage{amsfonts}
\usepackage{amsmath}
\usepackage{amsthm}
\usepackage{amssymb}
\usepackage{graphicx}
\usepackage{epspdfconversion}
\usepackage{cancel}
\usepackage{color}
\usepackage{caption}
\usepackage{subcaption}
\SectionNumbersOn

\author{Peter Poier}
\affiliation{Faculty of Physics, University of Vienna, Boltzmanngasse 5, A-1090 Vienna, Austria}
\email{peter.poier@univie.ac.at}
\author{Christos N.~Likos}
\affiliation{Faculty of Physics, University of Vienna, Boltzmanngasse 5, A-1090 Vienna, Austria}
\author{Angel J. Moreno}
\affiliation{Centro de F\'{i}sica de Materiales (CSIC-UPV/EHU) and Materials Physics Center MPC, Paseo Manuel de Lardizabal 5, E-20018 San Sebasti{\'a}n, Spain}
\alsoaffiliation{Donostia International Physics Center, Paseo Manuel de Lardizabal 4, E-20018 San Sebasti{\'a}n, Spain}
\author{Ronald Blaak}
\affiliation{Faculty of Physics, University of Vienna, Boltzmanngasse 5, A-1090 Vienna, Austria}

\title{An Anisotropic Effective Model for the Simulation of Semiflexible Ring Polymers}

\begin{document}

\begin{abstract} We derive and introduce anisotropic effective pair potentials to coarse-grain solutions of semiflexible rings polymers
of various lengths. The system has been recently investigated by means of full monomer-resolved computer simulations, revealing
a host of unusual features and structure formation, which, however, cannot be captured by a rotationally-averaged effective pair
potential between the rings' centers of mass [M. Bernabei {\it et al.}, Soft Matter {\bf 9}, 1287 (2013)]. 
Our new coarse-graining strategy is to picture each ring as a soft, penetrable disc.
We demonstrate that for the short- and intermediate-length rings the new model is quite capable of capturing the Physics in a 
quantitative fashion, whereas for the largest rings, which resemble flexible ones, it fails at high densities. Our work opens the way
for the physical justification of general, anisotropic penetrable interaction potentials. 
\end{abstract}

\section{Introduction}
\label{sec:introduction}
By the simple process of joining the ends of a linear polymer chain, one obtains a ring polymer (RP)\cite{micheletti2011polymers}. While the architecture of ring polymers is very simple, they differ in many interesting ways from their linear counterparts and are the subject of active research in Physics, Biology, Chemistry and even pure Mathematics. One interesting consequence is that for a dynamics that disallows strand crossing, there are different classes of configurations of a RP, which can never transform into each other. These are referred to as the \textit{topology classes} or \textit{knot types} of a RP. Knot theory is a fascinating and active branch of mathematics with many open problems concerning the enumeration and classification of knots\cite{adams1994knot}. A RP that has the topology of a circle is called an \textit{unknotted} RP.

Unlike RPs, linear polymer chains can strictly speaking never be knotted, as every configuration of a linear polymer chain can be continuously transformed to a straight line, without the need for strand crossings. While this is a fundamental difference between linear polymer chains and RPs, it is nevertheless possible to extend the concept of knots to \textit{physical} knots on linear polymer chains\cite{tubiana2011probing}. There are many works dealing with the properties of these \textit{physical} knots on linear polymer chains\cite{grosberg2007metastable,matthews2012effect,poier2014influence,dai2014metastable,di2014driving,renner2014untying,micheletti2014knotting,dai2015origin}, in particular due to their relevance in Biophysics, where they are for instance found in DNA\cite{sogo1999formation,arsuaga2002knotting} and can have significant effects on key processes\cite{portugal1996t7,deibler2007hin,liu2009and}.

The topological constraint of a ring polymer has important consequences for its physical behavior. It took the work of many authors\cite{des1981polymers,grosberg1996flory,deutsch1999equilibrium,grosberg2000critical,dobay2003scaling,moore2004topologically,moore2005limits,mansfield2010properties} to establish that the diameter of gyration $\rm D_g$ of an isolated ideal RP with fixed topology scales as $\langle {\rm D_g^2}\rangle \sim N^{2\nu}$, where $\nu\approx 0.588$ is the Flory exponent, which also describes the scaling behavior of the radius of gyration of self-avoiding linear chains\cite{de1979scaling}. This is true for all knot types, even for an ideal unknotted RP (i.e., without monomer excluded volume interactions, just keeping the topological constraints). An ideal linear polymer chain, on the other hand, remains more compact and exhibits the scaling law of a non-self-avoiding random walk $\langle {\rm D_g^2}\rangle \sim N$. Another important difference lies in the effective potential between the centers of mass of RPs. While the effective potential vanishes between infinitely thin linear polymer chains and also between an infinitely thin linear polymer chain and a RP, there remains a nonzero repulsive contribution between cyclic polymers with fixed topology\cite{bohn2010topological,narros2010influence}. Here one usually speaks of a topological potential. Furthermore, it was shown that the effective potential between two moderately sized RPs increases with the knot complexity of the rings\cite{narros2010influence}.

Also for concentrated systems, the topology of polymer chains plays an important role. The scaling of linear polymer chains in the melt is the one of a non-self-avoiding random walk $\langle {\rm D_g^2}\rangle \sim N$. Simulations of dense systems of RPs\cite{vettorel2009statistics,halverson2011molecularStatics} on the other hand showed that while short chains also exhibit a Gaussian scaling behavior $\langle {\rm D_g^2}\rangle \sim N$, long chains are compact and thus scale as $\langle {\rm D_g^2}\rangle \sim N^{2/3}$. In between there is a broad crossover region, where a $\langle {\rm D_g^2}\rangle \sim N^{4/5}$ scaling  provides a good description of the data. For the dynamics, it is expected that concatenations of ring polymers can have a significant effect, as they are permanent, in contrast to the entanglement of linear polymer chains. However, this implies that those concatenations are there in the first place, i.e., from the very synthesis of the sample on. Even in the absence of concatenations, there are important differences in the dynamics of RPs in the melt with respect to their linear counterparts. For instance recent experiments\cite{kapnistos2008unexpected} and simulations\cite{halverson2011molecularDynamics} revealed a power-law stress relaxation instead of the rubbery plateau found for linear chains.

For the large intermediate density domain between dilute solutions and melts, there are relatively few theoretical results despite the practical relevance of this regime for instance in the field of biophysics, where the topological interactions between chromatin loops plays a crucial role in the creation of chromosome territories\cite{bohn2010topological,marenduzzo2009topological,dorier2009topological,marenduzzo2010biopolymer}. A fruitful and modern approach for the economic description and simulation of macromolecules in this regime is the method of coarse-graining. The idea behind this method is to bridge the time and length scales in the system by describing the macromolecules via an effective model with a reduced set of suitably-chosen effective degrees of freedom (d.o.f.). The microscopic information of the monomer-resolved model is underlying the effective model, as it determines the form of the effective potential, which describes the interaction between the macromolecules. The advantage of this method is not only that every timestep in a simulation requires less computational effort due to the simplified representation, but also that one can often choose a much larger timestep in a simulation of the coarse grained model, as the d.o.f. that remain in the coarse-grained model change much slower in time than their counterparts in the monomer-resolved model\cite{likos2001effective,likos2006soft}.

The method of coarse-graining is well-established and has for instance found successful application for polymer chains\cite{louis2000can,krakoviack2003influence,narros2014multi}, star polymers\cite{likos1998star,jusufi2009colloquium,marzi2012coarse,capone2012telechelic}, star-shaped polyelectrolytes\cite{jusufi2002counterion,huissmann2009star}, dendrimers\cite{ballauff2004dendrimers,gotze2004tunable,huissmann2011effective}, and block copolymers\cite{pierleoni2006multiscale,pierleoni2007soft,capone2008entropic}. The identification of the relevant degrees of freedom is an essential part in the design of an effective model. One often uses isotropic effective models, where the macromolecules consisting of many individual monomers are reduced to their center of mass. For the semiflexible ring polymers such a model has already been investigated in ref.\cite{bernabei2013fluids}. Clustering was observed in monomer-resolved simulations of semiflexible ring polymers, as well as in the corresponding isotropic effective model. However, it was also shown that the monomer-resolved system shows anisotropic features that can not be accounted for in the isotropic effective model. Also the correlation functions stemming from the isotropic effective model are markedly different from the microscopically derived ones. Anisotropy is particularly strong for rings with high bending stiffness or few monomers, as they have a strong tendency to orient with respect to other rings in their proximity. This motivates us to introduce an anisotropic effective model for the description of semiflexible ring polymers in this article. In this model, we will define the effective particles as soft disc-like molecules which are described not only by their center of mass but also by the direction in which their faces are oriented. An anisotropic effective model was already used successfully for the description of hard disc-like macromolecules\cite{heinemann2014angle}, but to the best of our knowledge this approach was up to now never applied to penetrable macromolecules, where the centers of mass of the macroparticles can coincide. Penetrable particles are particularly interesting as they allow for clustering, which often leads to a rich phase behavior. For instance, point-particles interacting with a certain class of ultrasoft potentials form so-called cluster crystals\cite{mladek2006formation,lenz2012microscopically,sciortino2013computational}. Unlike in an ordinary crystal multiple soft particles can sit on top of each other at the same lattice site in a cluster crystal. Another peculiar feature of this state of matter is that by compressing it one only changes the occupation number of particles per lattice site, while the lattice constant remains invariant. Monomer-resolved simulations of semiflexible ring polymers on the other hand show the formation of the cluster glass phase\cite{coslovich2012cluster,Slimani2014}, which is an arrested-state that also contains some of the features found in the cluster crystal phase. In both cases the overall structure of the system is frozen, while individual particles can hop between the lattice sites of the cluster crystal or the stacks found in the cluster glass. Elongated dendrimers, which unlike hard rod-like particles exhibit local antinematic order\cite{georgiou2014}, are another interesting example for a a system of penetrable particles, which behaves distinctively different to its solid counterpart. By creating an anisotropic effective model for the semiflexible ring polymers, we aim at a model that is still computationally cheap, and that improves the description obtained by the isotropic model, especially for the case of high densities. In addition, the analysis of the interactions in the anisotropic effective potential allows us to get a better understanding for the interaction between the anisotropic, penetrable nanoparticles in systems of semiflexible RPs.

The remainder of this article is structured as follows. In Section \ref{sec:AnisoEffModel} we first present the Hamiltonian of the monomer-resolved model which we use for the description of semi-flexible RPs and then introduce an anisotropic effective model for such a system. In Section \ref{sec:SelfConsistentHisto} we give more details about the derivation of the effective interactions for such a model. We carried out Molecular Dynamics simulations of the monomer-resolved model and Monte Carlo simulations of the effective models; details concerning these simulations are given in Section \ref{sec:simulationDetails}. In Section \ref{potential:sec} we present the anisotropic effective potential
and discuss its features. Results of Monte Carlo simulations with this potential, which show that the inclusion of anisotropy in the effective model can significantly improve the agreement with the monomer-resolved model, are presented in Section \ref{manybody:sec}, whereas in Section \ref{truncate:sec} we briefly discuss the effects of truncation of the expansion of the potential on the quality of the results. Conclusions are given in Section \ref{sec:conclusions}. In the Appendix, we explain the expansion of the anisotropic pair-correlation function of a system of two RPs, which contains all the information for calculating the effective potential, as a sum of suitably chosen basis functions.

\section{Anisotropic Effective Model}
\label{sec:AnisoEffModel}
\subsection{Monomer-resolved Model for Semiflexible Ring Polymers} 
The derivation of the anisotropic effective model is based on a microscopic model of semiflexible ring polymers, each consisting of $N$ monomers. They are described with the bead-spring model by Kremer and Grest\cite{kremer1990dynamics} and an additional rigidity term. Thus, any two-monomers interact via the truncated and shifted Lennard-Jones potential
\begin{eqnarray}
{\rm V_{LJ}}(r)= \left\{
     \begin{array}{ll}
       4 \epsilon \left[\left(\frac{\sigma}{r}\right)^{12}-\left(\frac{\sigma}{r}\right)^{6}+\frac{1}{4} \right]  & \text{if } r< 2^{1/6}\sigma;\\
       0 & \text{if } r\geq 2^{1/6}\sigma.
     \end{array} \right.
\label{VLJ}
\end{eqnarray}
This potential is purely repulsive, accounting then for monomer excluded volume interactions. Bonded monomers also interact through a finitely extensible non-linear elastic potential (FENE)
\begin{eqnarray}
{\rm V_{FENE}}(r)=-\frac{k R_0^2}{2} \ln \left[ 1- \left(\frac{r}{R_0} \right)^2\right].
\label{VFENE}
\end{eqnarray}
Rigidity is introduced via the bending potential
\begin{eqnarray}
{\rm V_{bend}}(\theta)= \kappa (1-\cos\theta)^2, 
\label{VBend}
\end{eqnarray}
where $\theta$ is the angle between two consecutive bond vectors.\footnote{Note that this is not the Kratky-Porod model (linear in the cosine). We expect the same qualitative results for Kratky-Porod rings with the same $N$'s and persistence lengths as the model simulated here"}. The potential ${\rm V_{bend}}$ vanishes for $\theta=0$, when the polymer chain does not bend at the respective angle. We choose $\epsilon= k_B T$, $k=30 k_B T/\sigma^2$, $R_0=1.5 \sigma$ and $\kappa= 30 k_B T$, where $k_B$ is the Boltzmann constant and $T$ the temperature. These are precisely the parameters employed in the simulation study of ref. \cite{bernabei2013fluids}. The corresponding dynamics does not allow for chain crossings and thus topology is preserved. 

In ref.\cite{Slimani2014} the characteristic ratio\cite{rubinstein2003polymer} of this polymer model was estimated by carrying out simulations of isolated linear chains. Excluded volume interactions were switched off except for mutually connected monomers, in order to obtain long-range Gaussian statistics. $C_\infty$ was obtained by analyzing the long-$s$ limit of the ratio $\langle R^2(s)\rangle /s\langle b^2\rangle$,
where $R(s)$ is the distance between two monomers $i,j$ with $s = |i-j|$, and $b$ is the bond length ($\langle b^2\rangle = 0.94$). The authors reported a value of $C_\infty\sim 15$, which is typical for stiff polymers\cite{rubinstein2003polymer}. We can give an estimate for the persistence length of the model by mapping it to the freely rotating chain model using the relation $\cos\theta= \frac{C_\infty-1}{C_\infty+1}\sim0.875$, where $\theta$ is the bending angle of the freely rotating chain model\cite{rubinstein2003polymer}. The persistence length is then obtained as $s_p b=-b/\ln(\cos\theta)\sim 7.3$. We carried out simulations of ring polymers with $N= 20$, $50$, and $100$ monomers, which have the contour to persistence length ratio $N/s_p\sim 2.7$, $6.7$ and $13.3$ respectively.

\subsection{The Anisotropic Effective Model}

In earlier work\cite{bernabei2013fluids}, Bernabei {\it et al.} carried out extensive monomer-resolved simulations of stiff ring polymers to obtain the structure
of concentrated solutions of the same. In an attempt to coarse-grain the system in the simplest possible way, they also derived and  
employed an isotropic effective potential for their effective description,
reducing thereby stiff RP's into point-like effective particles, namely their centers of mass.  
At this level of approximation, the effective particles possess no other, internal (spin-like) degrees of freedom and thus the effective interaction is isotropic. 
The effective potential between these macroparticles was defined by calculating the pair correlation function ${\rm g^{iso}}(r)$ in an infinitely dilute system and using
\begin{eqnarray}
{\rm \beta V^{iso}_{eff}}(r)\equiv-\ln \left[{\rm g^{iso}}(r)\right]
\label{eq:VEffIso}
\end{eqnarray}
to define the effective interaction potential between these point particles, where $\beta=1/(k_BT)$. In the infinitely dilute case, the distribution of the centers of mass of the ring polymers in equilibrium is identical to the distribution of the point particles in the effective model. At higher densities, however, it turns out that multi-particle terms in the effective potential are necessary to obtain the correct equilibrium distribution of the centers of mass in the effective model. One of the reasons due to which multi-particle interactions become important is that two ring polymers that are sufficiently close and stiff, will prefer to align parallel to each other. A third ring polymer interacting with those two will not see them as two independent rings but as a system of two rings that are correlated. When using the potential ${\rm V_\text{eff}}(r)$ calculated in Eq. (\ref{eq:VEffIso}) one assumes that the free energy penalty of one ring with respect to a second is independent of the presence of another polymer in the vicinity of the second. If the ring polymers preferentially align parallel, this assumption is clearly violated and one has to correct the effective potential by introducing multi-particle terms. Bernabei {\it et al.} showed that already at moderate densities one can encounter strong correlations of the orientations of the semiflexible ring polymers, in particular if the chains contain only few monomers (e.g., $N=20$)\cite{bernabei2013fluids}. Therefore, a more accurate coarse-graining which takes into account the ring anisotropy is called for.

The easiest way to incorporate the correlations of the orientations of rings is to introduce them via additional degrees of freedom (d.o.f.) in the effective description. This is precisely what we do in this article. For this purpose, we need to first come up with a suitable definition for the orientation of a RP. To this end, we make use of the gyration tensor
\begin{eqnarray}
S_{\alpha\beta}=\frac{1}{N}\sum_{i=1}^N r_\alpha^{(i)} r_\beta^{(i)},
\end{eqnarray}
where $r_\alpha^{(i)}$ ($\alpha=x,y,z$, Cartesian components) denotes the position of the i-th monomer with respect to the center of mass of the ring to which this monomer belongs. The eigenvectors of $S_{\alpha\beta}$ are the principal axes of an ellipsoid that approximates the shape of the macromolecule: If a RP is flat, which means that all its monomers lie in one plane, the ellipsoid has one zero eigenvalue with a corresponding eigenvector that is perpendicular to that plane. Also in the more general case, where the monomers do not all lie in the same plane, we define the normalized eigenvector corresponding to the smallest eigenvalue of $S_{\alpha\beta}$ as the direction vector $\mathbf{d}$ of the RP. Note that $\mathbf{d}$ and $-\mathbf{d}$ are equivalent for reasons of symmetry. The ring polymers in the anisotropic effective model we propose are described via the position vectors of their centers of mass, $\mathbf{R}^{(i)}$, and their direction vector $\mathbf{d}^{(i)}$. Henceforth, we describe the stiff rings as soft circular discs and ignore differences in the other 2 eigenvectors of the gyration tensor $S_{\alpha\beta}$. This choice is motivated by the limit of infinite bending stiffness, where the rings assume flat and precisely circular conformations. Our model, therefore, amounts to the {\it minimal} anisotropic extension of the spherically symmetric effective interaction between the centers of mass. We emphasize, however, that there is no a priori guarantee that this will be an improvement over the isotropic model at finite densities and in particular at high concentrations: this depends on the degree in which the RP's at high concentration maintain their anisotropic shape and properties encoded in the high-dilution limit in which the anisotropic pair potential is derived. Accordingly, the introduction of such a potential is not a straightforward part of a systematic strategy of introducing more and more detail into the effective description of the system. 

In order to determine the anisotropic effective potential ${\rm V_{eff}}$, we carried out monomer-resolved simulations of two ring polymers inside a large simulation box. The effective pair potential is then defined such that it exactly reproduces the correlation functions of the effective degrees of freedom in this infinitely dilute, monomer-resolved simulation. In the effective model, two ring polymers are described by a total of 10 degrees of freedom, three for each center of mass and two for each direction vector of each ring polymer. However, due to translation, rotation and mirror symmetry the distinct configurations (those that cannot be related by symmetry transformations) of a system with two effective particles are reduced and can be specified by 4 parameters only.

\begin{figure}[htp]
\includegraphics[width=12.0cm]{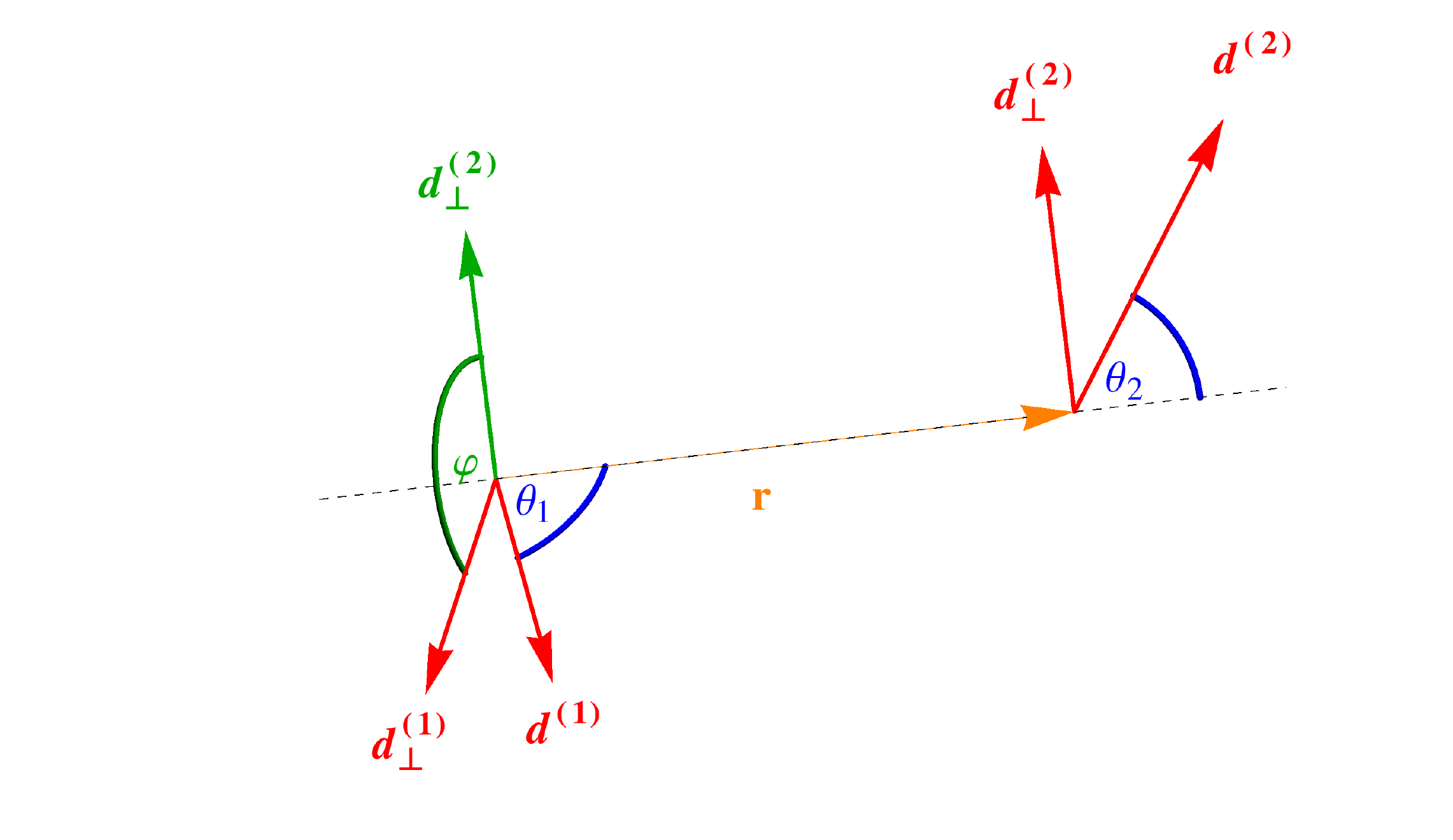} 
\caption{Illustration of the effective variables $r=|\mathbf{r}|$, $\theta_i$ and $\varphi$ with which relative configurations of ring polymers are described.}
\label{anglesScetch}
\end{figure}

A convenient choice for these variables is illustrated in figure \ref{anglesScetch} and reads as follows:
\begin{eqnarray}
r &\equiv& |\mathbf{r}|;\nonumber\\
\cos\theta_1 & \equiv & \mathbf{d}^{(1)} \cdot \hat{\mathbf{r}};\nonumber\\
\cos\theta_2 & \equiv & \mathbf{d}^{(2)} \cdot \hat{\mathbf{r}};\nonumber\\
\varphi & \equiv &\arccos \left(\frac{\mathbf{d}^{(1)}_\perp\cdot\mathbf{d}^{(2)}_\perp}{|\mathbf{d}^{(1)}_\perp| |\mathbf{d}^{(2)}_\perp|}\right),
\label{eq:effCoordsDef}
\end{eqnarray}
where $\mathbf{r}\equiv\mathbf{R}^{(2)}- \mathbf{R}^{(1)}$ is the connection vector between the centers of mass of the two rings, $\hat{\mathbf{r}}\equiv \mathbf{r}/r$ the unit vector in the direction of $\mathbf{r}$ and  $\mathbf{d}^{(i)}_ \perp$ the component of the director $\mathbf{d}^{(i)}$ perpendicular to $\mathbf{r}$; $0\leq\varphi\leq\pi$ denotes the angle between vectors $\mathbf{d}^{(1)}_\perp$ and $\mathbf{d}^{(2)}_\perp$. By selecting the appropriate sign of $\mathbf{d}^{(i)}$ we can always choose $\cos\theta_i$ to lie in the interval $[0,1]$.

Let us define the ideal case as the system where the effective particles do not interact and thus every orientation and position of the effective particles occurs with equal probability, independently of the configuration of the other effective particle. With the effective coordinates defined in (\ref{eq:effCoordsDef}) the probability density in the ideal system, ${\rm P_{id}}\left(r,\cos\theta_1,\cos\theta_2,\varphi\right)$ is proportional to $r^2$ and constant in both $\cos\theta_i$ as well as in $\varphi$. This simple behavior of ${\rm P_{id}}\left(r,\cos\theta_1,\cos\theta_2,\varphi\right)$, makes (\ref{eq:effCoordsDef}) a particularly convenient choice of the effective coordinates.
In the simulations with two ring polymers, we obtain a probability density ${\rm P}\left(r,\cos\theta_1,\cos\theta_2,\varphi\right)$ that is different from the ideal distribution ${\rm P_{id}}$. We define as a generalised version of the radial distribution function the anisotropic pair correlation function as
\begin{eqnarray}
{\rm g}\left(r,\cos\theta_1,\cos\theta_2,\varphi\right)={\frac{{\rm P}\left(r,\cos\theta_1,\cos\theta_2,\varphi\right)}{{\rm P_{id}}\left(r,\cos\theta_1,\cos\theta_2,\varphi\right)}}.
\label{eq:gPRelation}
\end{eqnarray}
Thus, the quantity ${\rm g}\left(r,\cos\theta_1,\cos\theta_2,\varphi\right)$ describes the factor by which configurations in the effective anisotropic model have to be enhanced or suppressed with respect to the ideal case, in order to obtain a distribution for the effective d.o.f. that is identical to the distribution obtained in a monomer-resolved simulation in the infinitely dilute case. As in the isotropic case (\ref{eq:VEffIso}) the relation to the associated effective potential reads
\begin{eqnarray}
\beta{\rm V_{eff}}\left(r,\cos\theta_1,\cos\theta_2,\varphi\right)=-\ln\left[g\left(r,\cos\theta_1,\cos\theta_2,\varphi\right)\right].
\end{eqnarray}
From the anisotropic effective potential, we can deduce the isotropic pair-correlation function via
\begin{eqnarray}
\label{eq:gIsoGRelation}
{\rm g^{iso}}(r)=\frac{1}{\pi}\int_{0}^{1} {\rm d}\cos\theta_1\int_{0}^{1}{\rm d}\cos\theta_2\int_0^{\pi}{\rm d}\varphi\;{\rm g}\left(r,\cos\theta_1,\cos\theta_2,\varphi\right).
\end{eqnarray}
The associated isotropic effective potential is then given by (\ref{eq:VEffIso}).

The effective potential between two identical ring polymers remains invariant if we swap the orientations of their respective director with respect to the connection vector, i.e., if we swap the values of the polar angles $\theta_1$ and $\theta_2$ of the two rings:
\begin{eqnarray}
{\rm V_{eff}}\left(r,\cos\theta_1,\cos\theta_2,\varphi\right)= {\rm V_{eff}}\left(r,\cos\theta_2,\cos\theta_1,\varphi\right).
\end{eqnarray}
This symmetry is violated for the effective potential between bidisperse ring polymers, e.g. for ring polymers with a different number of monomers. However, apart from this symmetry, there would be no differences in the procedure of calculating the effective potential between different types of ring polymers.

\section{Simulation Details}
\label{sec:simulationDetails}
\subsection{Derivation of the effective interaction}
\label{subsec:derivEffInt}
To determine the effective potential, we carry out constant $NVT$ molecular dynamics simulations with two ring polymers. We simulate rings of $N=20$, $50$ and $100$ monomers. For these simulations we use the LAMMPS simulation package\cite{plimpton1995fast}. The polymer rings are placed in a simulation box, which is large enough to prevent multiple interactions via the periodic boundary conditions. The temperature in the simulation is maintained by the use of Langevin-Dynamics. The corresponding equations of motion read as\cite{allen1989computer}:
\begin{eqnarray}
m \mathbf{\ddot{r}}_i(t)= \mathbf{F}_i(t)- \gamma m \mathbf{\dot{r}}_i(t)+\boldsymbol{\eta}_i(t).
\end{eqnarray}
Here, $\mathbf{r}_i$ is the position of the $i$-th monomer, $m$ its mass and $\mathbf{F}_i(t)$ the deterministic force acting on it, which includes the microscopic forces originating from potentials (\ref{VLJ}-\ref{VBend}) and the force originating from a biasing potential. The bias potential ${\rm V_{bias}}=\left(r-r_j\right)^2 k_j/2$ introduces a harmonic spring with spring constant $k_j$ between the centers of mass of the ring polymers. The spring is relaxed for $r=r_j$. We carried out simulations for different values of $r_j$, starting at $r_j=0$ and increasing it up to some maximum value $r_C$ in steps of $\sigma/2$. For $N=20$, $50$, $100$, $r_C$ was chosen as $10\sigma$, $20\sigma$, $30\sigma$ respectively. These values for $r_c$ are much bigger than the infinite-dilution diameters of gyration (${\rm D_{g0}}=$ $5.9\sigma$, $13\sigma$, $21.5\sigma$ for $N=$ 20, 50, 100 respectively). For $k_j$ we chose the values $2.5 \epsilon/\sigma^2$ and $\epsilon/\sigma^2$ for all ring sizes and for $N=20$ we also carried out simulations with $k_j= 5 \epsilon/\sigma^2$. The quantity $\boldsymbol{\eta}_i(t)$ is a random force, with $\langle\boldsymbol{\eta}_i(t)\rangle=0$,
which is related to the friction coefficient $\gamma$ by the fluctuation dissipation relation $\langle \eta_i^\alpha(t) \eta_j^\beta(t') \rangle = 2 \gamma m k_B T \delta_{ij}\delta_{\alpha \beta}\delta(t-t')$, $\alpha$ and $\beta$ denoting Cartesian components.
Our unit of time is set by $t_0=(m \sigma^2/\epsilon)^{1/2}$ and the friction coefficient $\gamma$ is chosen as $1/t_0$. We integrate the equations of motion with a timestep of $\Delta t=10^{-3}t_0$, and use $2 \times 10^8$ timesteps for equilibrating the system and collect data during another $2 \times 10^9$ timesteps.

We sample histograms $P^{(j)}(Q)$, where $Q$ refers to a bin in the 4D space of the effective coordinates. $P^{(j)}(Q)$ gives the probability for a state in the $j$-th simulation to have effective coordinates in bin $Q$. The histograms have $128$ bins in $r$ and $16$ bins in $\cos \theta_1$, $\cos \theta_2$ and $\varphi$ direction. As discussed in appendix \ref{sec:SelfConsistentHisto} we use the \textit{Self-Consistent Histogram Method} by Ferrenberg and Swendsen\cite{ferrenberg1989optimized,frenkel2001understanding} to combine the $P^{(j)}(Q)$ histograms for simulations with identical ring sizes but different biasing potentials to arrive at an estimate for $P(Q)$ in the unbiased system.

\subsection{Many-body effective fluid}
Using the anisotropic effective potential we carry out standard Metropolis Monte Carlo (MC) simulations for the anisotropic effective model. The values of the anisotropic effective potential have been calculated on a discrete grid in the $(r,\cos \theta_1, \cos \theta_2, \varphi)$ space and we use linear interpolation to estimate the values of $\exp(-\beta {\rm V_{eff}})$ in between the grid points. Having in mind a comparison with both the monomer-resolved simulation results and the isotropic effective potential of ref.\cite{bernabei2013fluids}, we choose the same number of particles and effective densities that were used in those simulations. For rings with $N=20$, $50$ and $100$ monomers we simulate systems of $n=2400$, $1600$ and $1200$ rings, respectively, varying in each case accordingly the cubic box size $L$ as to achieve the desired density $\rho = n/L^3$. As the effective potential ${\rm V_{eff}}$ is bounded, a random distribution of the particles in the simulation box can be used as initial condition. We have implemented two types of MC moves: the first one translates a randomly chosen particle in a random direction, and the second randomly rotates the particle's director by some angle. The distance by which the particles are displaced and the angle by which they are rotated is randomly selected in an interval starting at $0$ and going up to some maximum value. For both moves, this maximum value of the interval is chosen such that the acceptance ratio is approximately $15\%$. We use $9\times10^6$ MC moves to equilibrate the system. During this equilibration period the individual soft particles diffuse to several times their own diameter. Afterwards, during $15\times10^6$ MC moves equilibrium configurations are generated. We store the configurations every $20 \times 10^3$ moves and use them to compute the physical observables that are presented in the section \ref{manybody:sec}.

The gain in computational efficiency for the simulations in going from a monomer resolved to a coarse grained simulation is considerably. The relevant quantity to consider here, would be the velocity through phase-space. This can conveniently be characterized by means of the mean square displacement of the rings per unit of CPU time. If one ignores the detailed implementation aspects, the CPU time spend on a single sweep over all monomers in the former and a run over all effective ring particles in the latter, which strictly speaking depend on both the number of monomers $N$ per ring and the overall density of rings, are of similar magnitude. However, the diffusion per sweep in the coarse grained simulation is significantly larger than that for the monomer resolved simulation, i.e., for the case of $N=50$ and $\rho^*=20$ this results in a factor of approximately $10^4$. The reason for this dramatic improvement is two-fold. First of all the translation/rotation of an effective ring corresponds to a much more time-consuming collective movement of the constituents. The second even more important contribution arises from the steric interaction that are present in the monomer resolved simulations and prevent the unphysical crossing of chain segments. In the coarse grained simulations such a restriction is absent, i.e., the effective rings are penetrable and can move apparently through each other. On this level of description this effect is not an unphysical process, but should be interpreted as a short-cut connecting initial and final configurations that are connected by a much more time-consuming and physically realizable pathway of folding and collective monomer movements.

\section{The Anisotropic Effective Potential}
\label{potential:sec}
\begin{figure}[htp]
\begin{center}
\includegraphics[width=12.0cm]{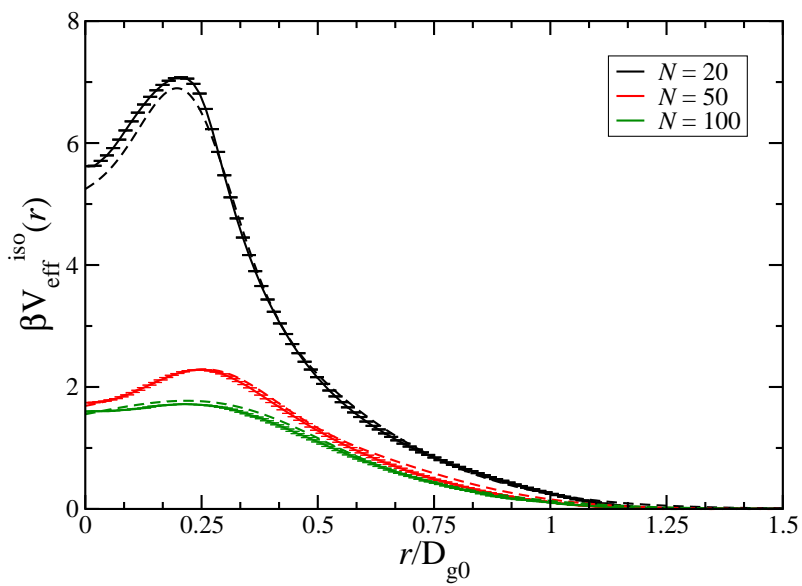} 
\end{center}
\caption{The effective, center-of-mass pair potential in the isotropic effective model for rings with different numbers of monomers $N$. 
The center-of-mass separation $r$ is scaled with $\rm D_{g0}$, the average diameter of gyration of a free ring polymer. The solid line shows the angularly-averaged effective pair potential, while the dashed lines are results of ref. \cite{bernabei2013fluids}.}
\label{vEffIsoPlot}
\end{figure}

We commence by recalculating the {\it isotropic}, i.e., angularly-average effective pair potential ${\rm V^{iso}_{eff}}(r)$ between the stiff rings, as a way of comparison
with the previously derived results in ref. \cite{bernabei2013fluids}. Results are summarized in 
figure \ref{vEffIsoPlot}, reproducing indeed the previously derived ones\cite{bernabei2013fluids}. The potential for $r=0$ is finite, as the rings are allowed to overlap. It also features a local minimum there, whereas its maximum is located at $r\approx 0.25 {\rm D_{g0}}$ for all ring types investigated. Here, $\rm D_{g0}$ is the average diameter of gyration of a free ring polymer. The height of the potential barrier at small distances of $r$ decreases by increasing the number of monomers $N$ on the ring polymers.

\begin{figure}[htp]
\begin{center}

\begin{subfigure}[t]{7.5cm}
\includegraphics[width=7.5cm]{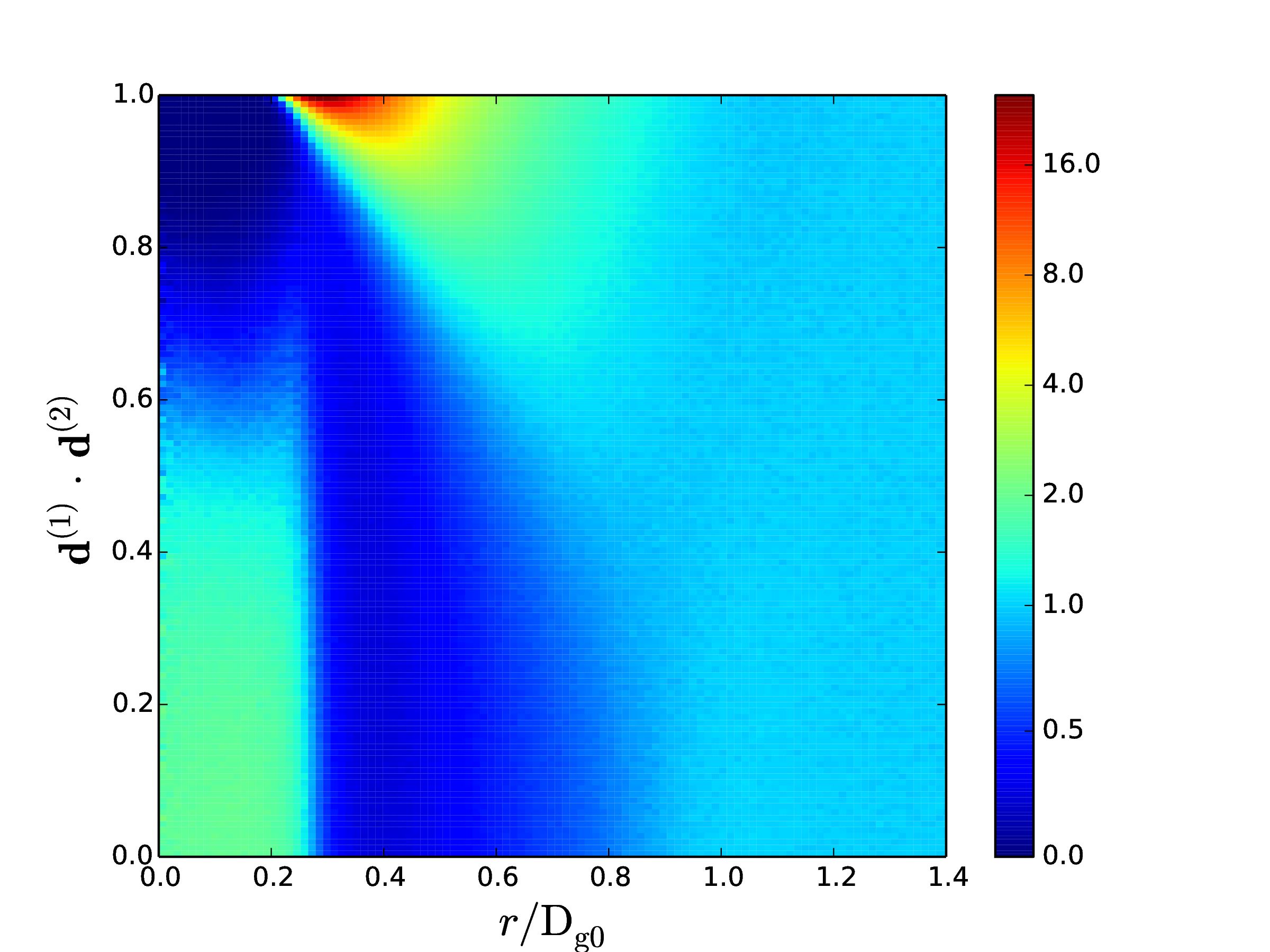}
\caption{(a) $N=20$}
\end{subfigure}
\begin{subfigure}[t]{7.5cm}
\includegraphics[width=7.5cm]{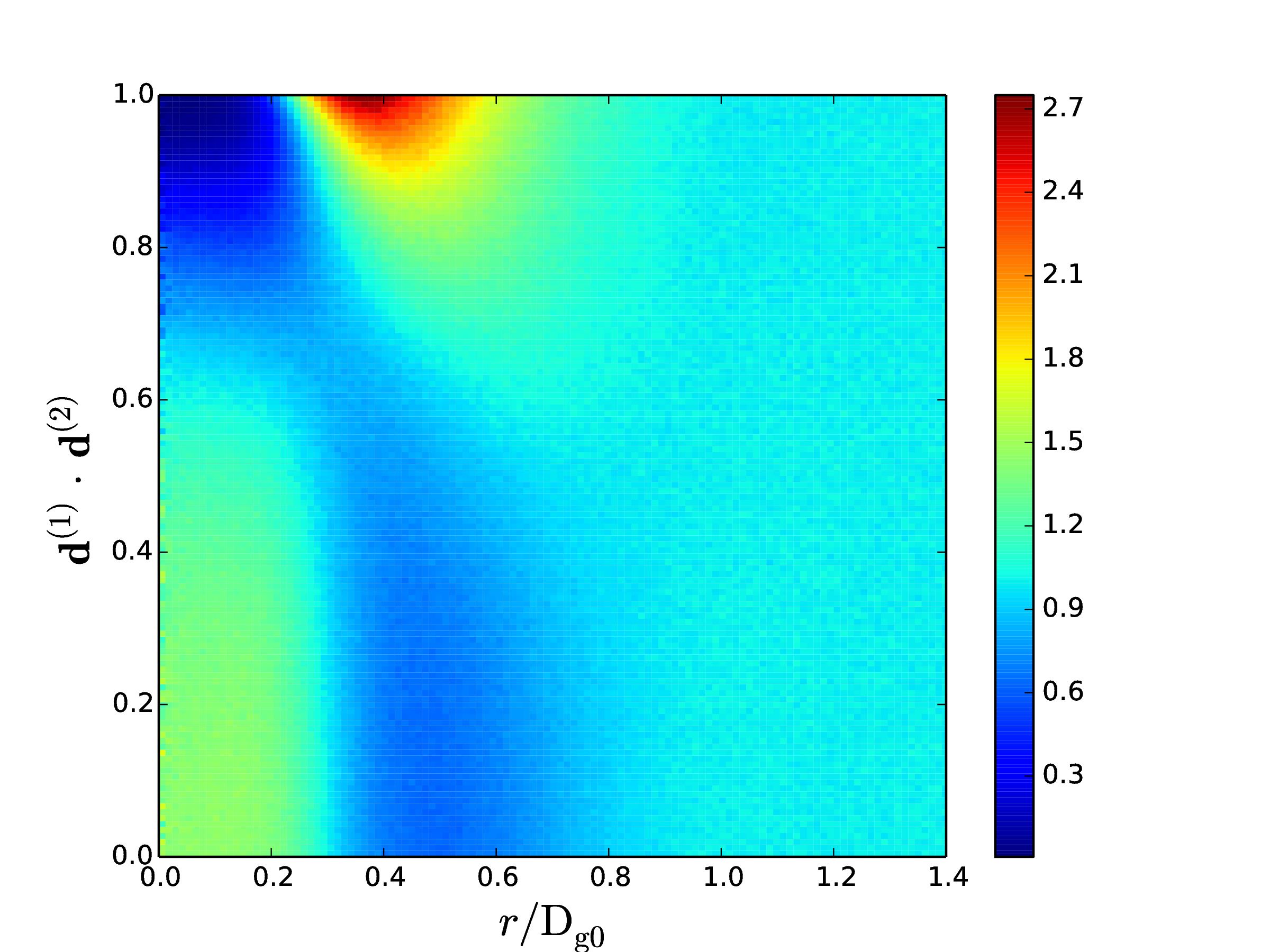} 
\caption{(b) $N=50$}
\end{subfigure}
\vspace{0.3cm}
\begin{subfigure}[t]{7.5cm}
\includegraphics[width=7.5cm]{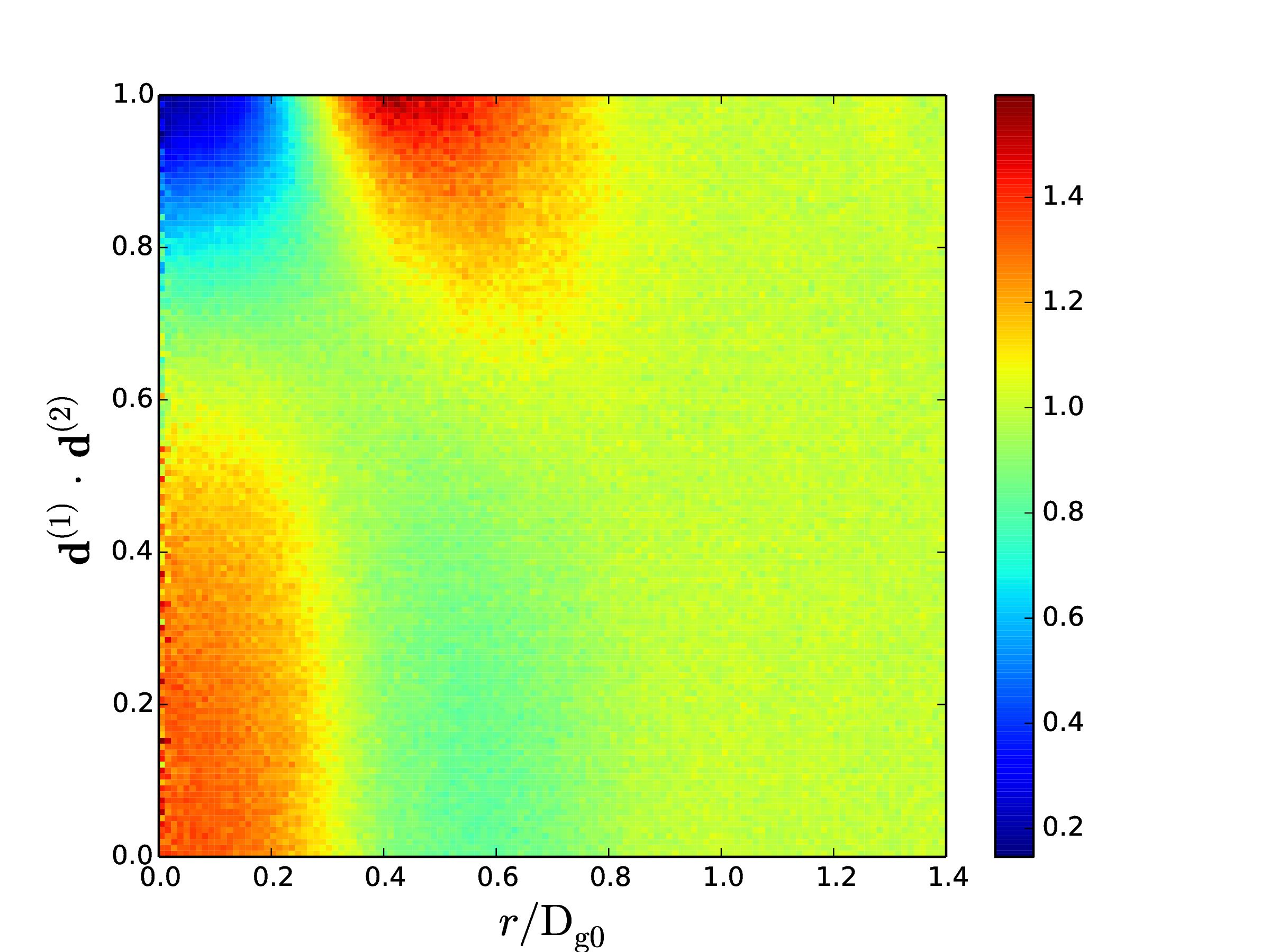} 
\caption{(c) $N=100$}
\end{subfigure}

\end{center}
\caption{Infinite-dilution limit of the quantity ${\rm G}(r, \mathbf{d}^{(1)}\cdot \mathbf{d}^{(2)})$, which quantifies the distribution of the scalar product between directors for different values of $r$. We visualize this distribution for the ring sizes $N=20$, $N=50$ and $N=100$.}
\label{gD1DotD2Plots}
\end{figure}

Going over to the anisotropic effective model, we proceed with computing the aforementioned pair correlation function ${\rm g}(r,\cos \theta_1, \cos \theta_2, \varphi)$ for a system of two ring polymers. As the latter depends on 4 effective coordinates and is therefore difficult to visualize, we first introduce a reduced pair correlation function
 ${\rm g}(r, \mathbf{d}^{(1)}\cdot \mathbf{d}^{(2)})$, which expresses the relative joint probability density of observing the two ring polymers at a distance $r$ 
 and with the directors mutually oriented at the value given by their scalar product, $\mathbf{d}^{(1)}\cdot \mathbf{d}^{(2)}$, over the same quantity
 for noninteractive rings. Moreover, we introduce a reduced
 version of this function, ${\rm G}(r, \mathbf{d}^{(1)}\cdot \mathbf{d}^{(2)})$, by dividing over its isotropic counterpart, i.e.,
\begin{eqnarray}
{\rm G}(r, \mathbf{d}^{(1)}\cdot \mathbf{d}^{(2)}) = \frac{{\rm g}(r, \mathbf{d}^{(1)}\cdot \mathbf{d}^{(2)})}{{\rm g^{iso}}(r)}.
\label{ggg:eq}
\end{eqnarray}
Results are summarized in figure \ref{gD1DotD2Plots}.  
One can see that the angular distribution of the directors changes significantly for $r \approx 0.25 {\rm D_{g0}}$, which is approximately the position of the maximum of ${\rm V^{iso}_{eff}}(r)$. For $r<0.25 {\rm D_{g0}}$ the angle between the directors is biased towards $\pi/2$, while for $0.25  {\rm D_{g0}}<r<{\rm D_{g0}}$ they prefer to align parallel with respect to each other. The position of the maximum of ${\rm V^{iso}_{eff}}(r)$ coincides approximately with the distance $r$ where interpenetrated configurations of the rings become sub-dominant and where they are more likely to align parallel to each other. The transition between these two domains is particularly steep for $N=20$ and becomes smoother for rings with a larger number of monomers. When the rings interpenetrate each other, the distribution of angles between the directors is rather wide, while it gets narrow after the transition where the bias towards parallel alignments of the rings is very strong in particular for the smallest rings with $N=20$. It is readily visible from figure \ref{gD1DotD2Plots} that anisotropy is particularly important for smaller rings. For $r> {\rm D_{g0}}$, the distribution of the angle between the directors becomes flat, as the rings are then well separated and hence do not interact. Note that by definition, Eq.\ (\ref{ggg:eq}), 
the quantity ${\rm G}(r, \mathbf{d}^{(1)}\cdot \mathbf{d}^{(2)})$ is a normalized probability distribution for {\it fixed} $r$, 
and in figure \ref{gD1DotD2Plots} it is therefore meaningless to compare the plotted function at different $r$ values.

\begin{figure}[htp]
\begin{center}
\begin{tabular}{cc}

\begin{subfigure}[t]{8.5cm}
\includegraphics[width=8.5cm]{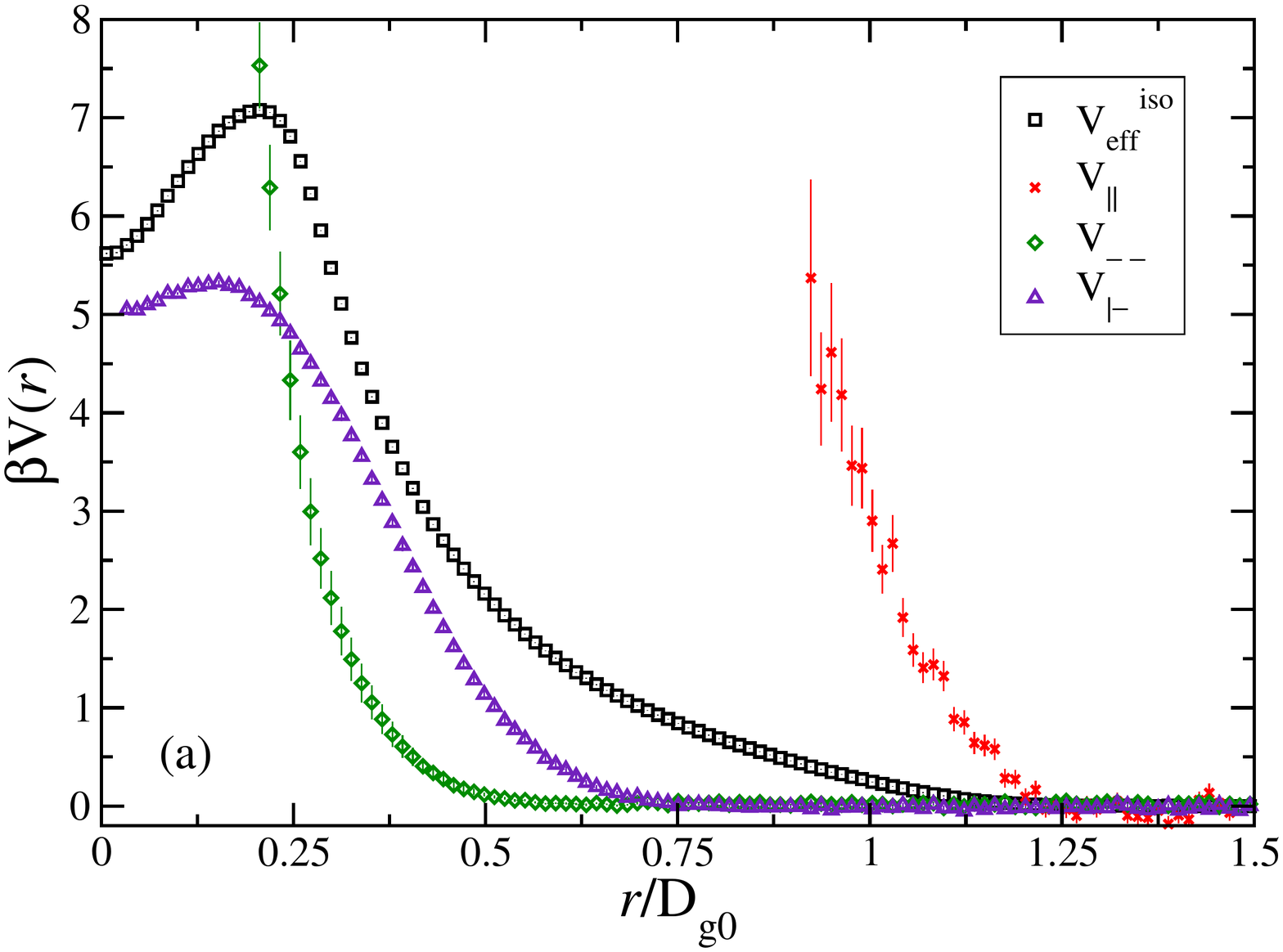}
\end{subfigure}
&
\begin{subfigure}[t]{8.5cm}
\includegraphics[width=8.5cm]{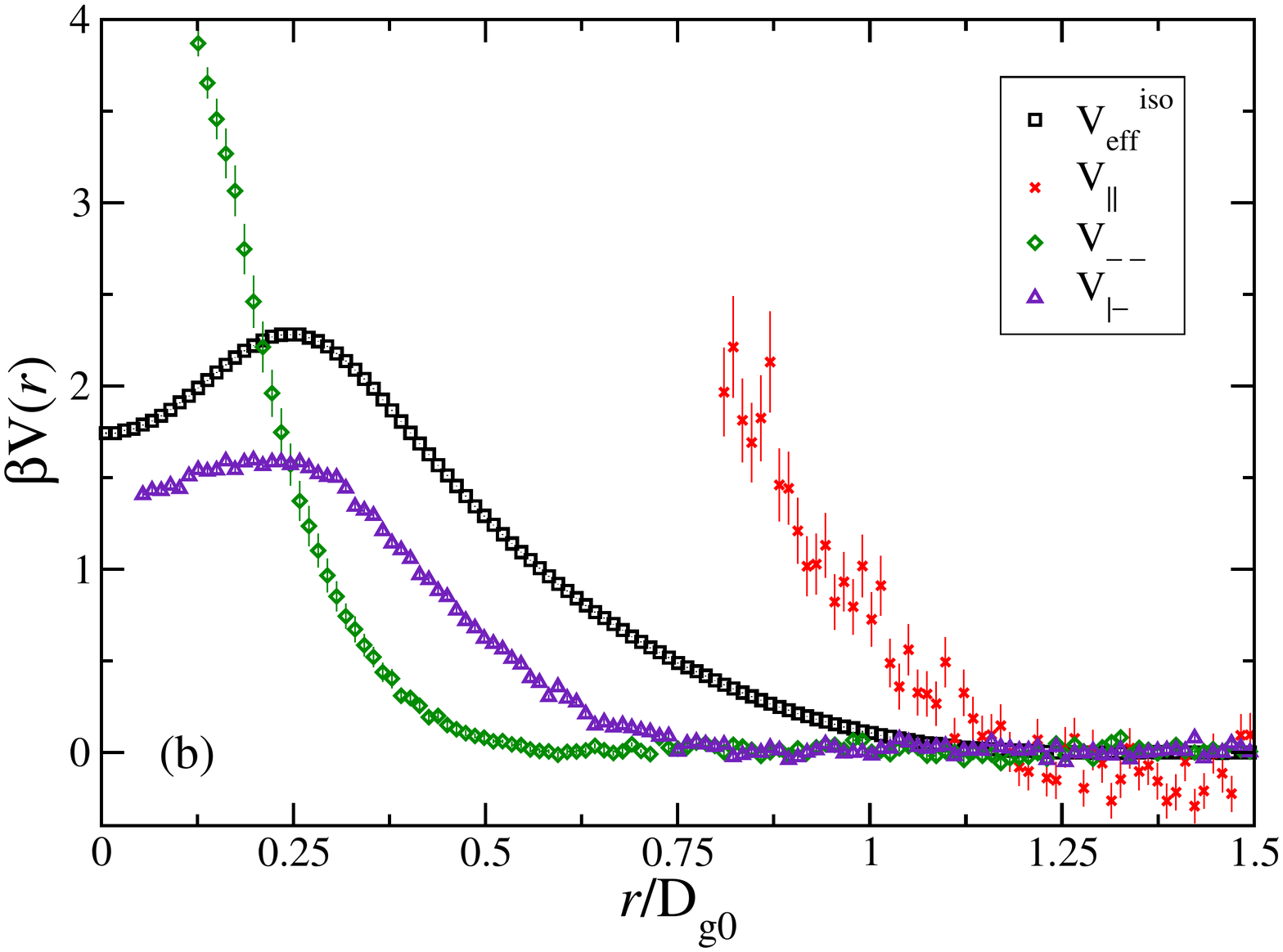} 
\end{subfigure}
\vspace{0.3cm}
\\
\begin{subfigure}[t]{8.5cm}
\includegraphics[width=8.5cm]{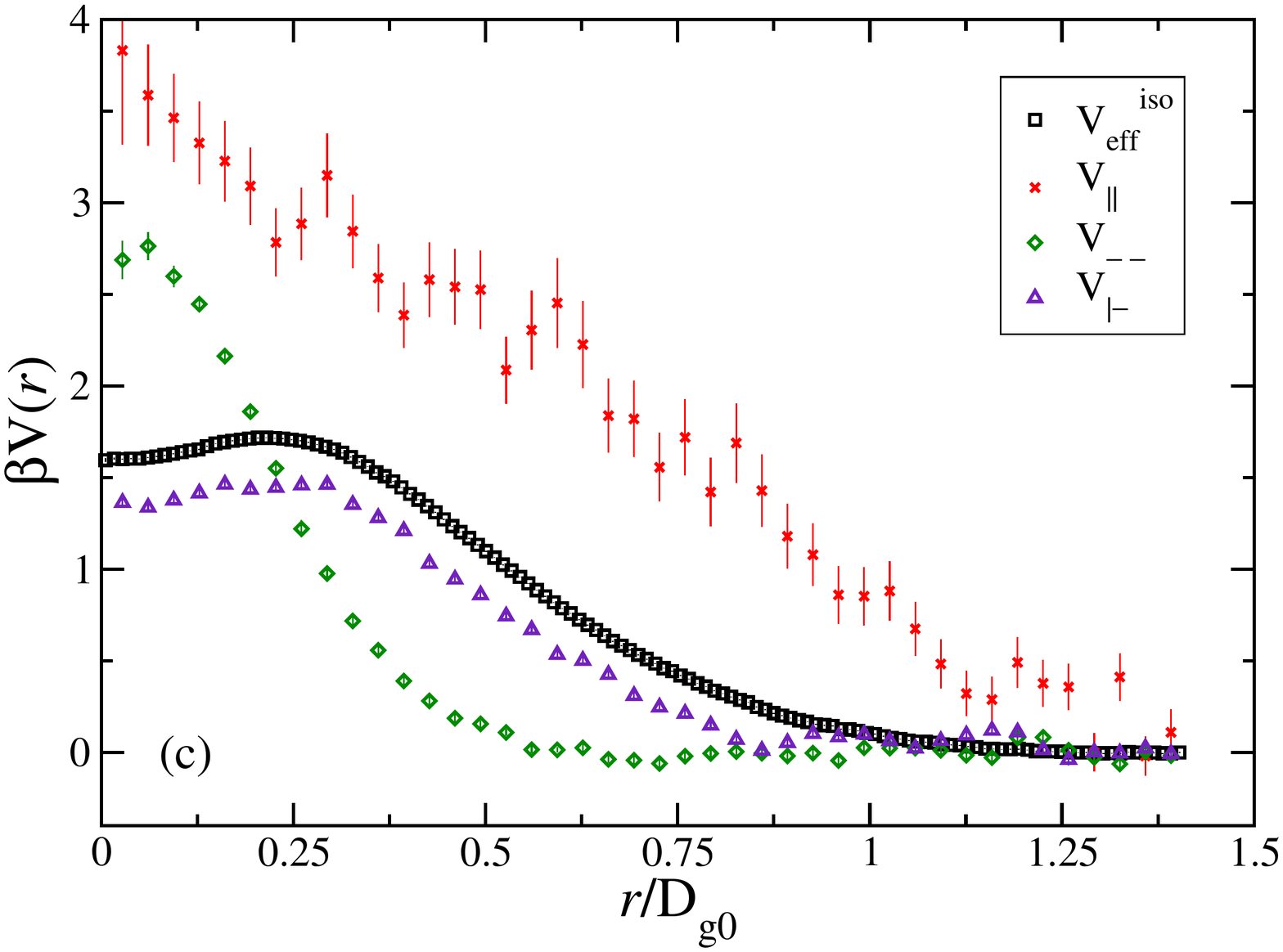} 
\end{subfigure}
&
\raisebox{0.7\height}{\begin{minipage}{6cm}
\centering
$||$= \raisebox{-0.4\height}{\includegraphics[width=4cm]{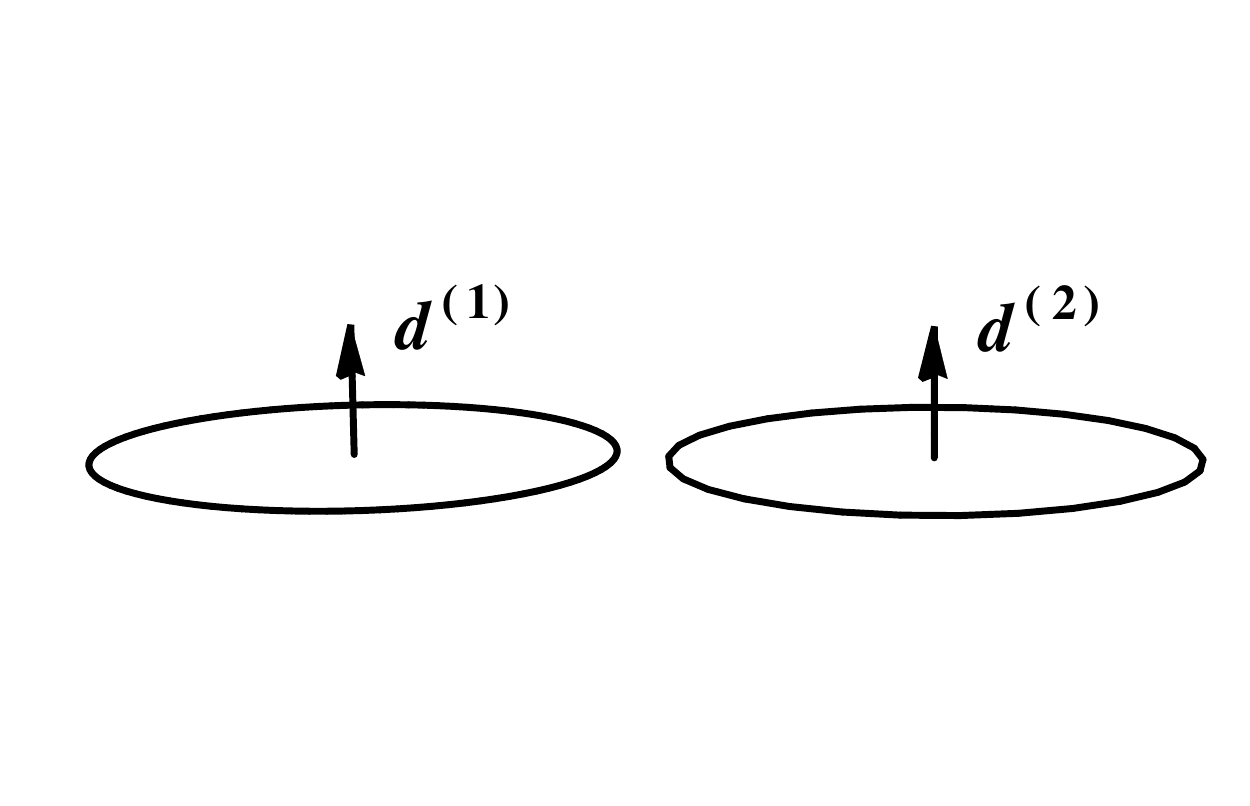}}
\\
$--$= \raisebox{-0.5\height}{\includegraphics[width=4cm]{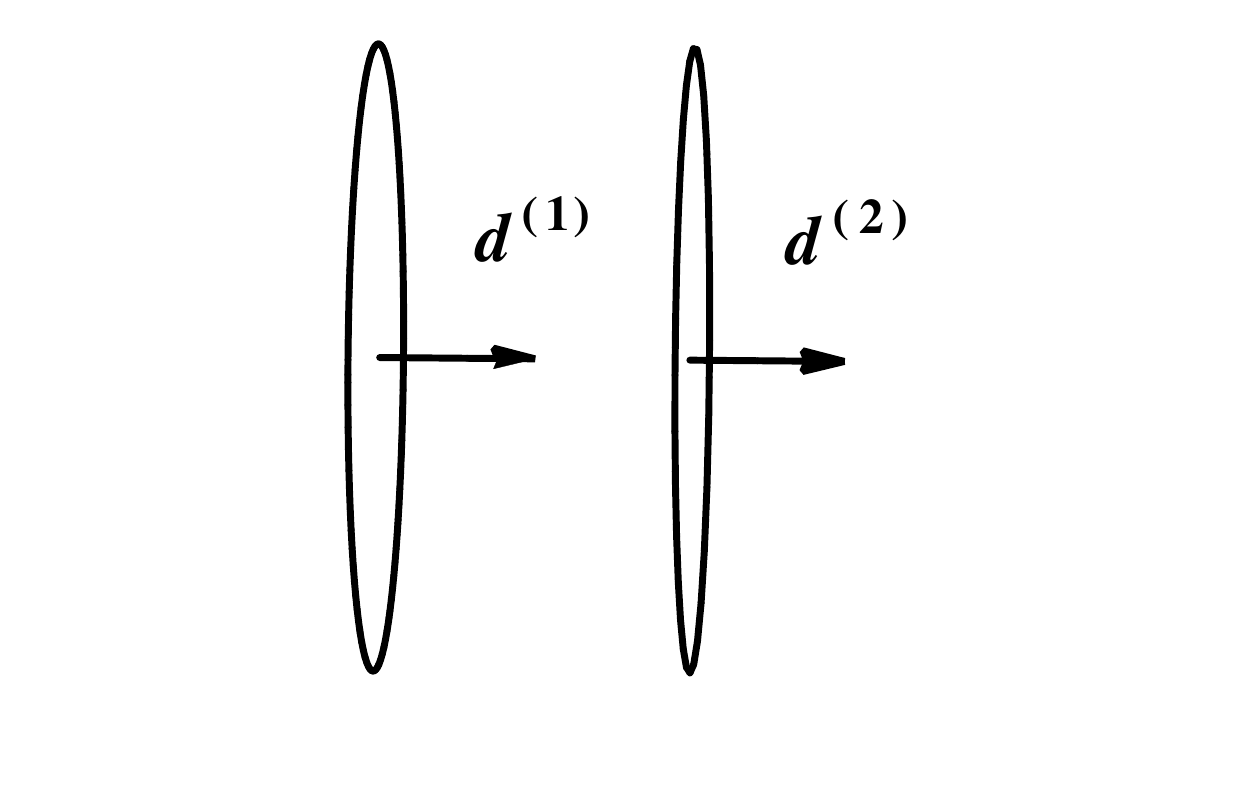}}
\\
$|-$= \raisebox{-0.5\height}{\includegraphics[width=4cm]{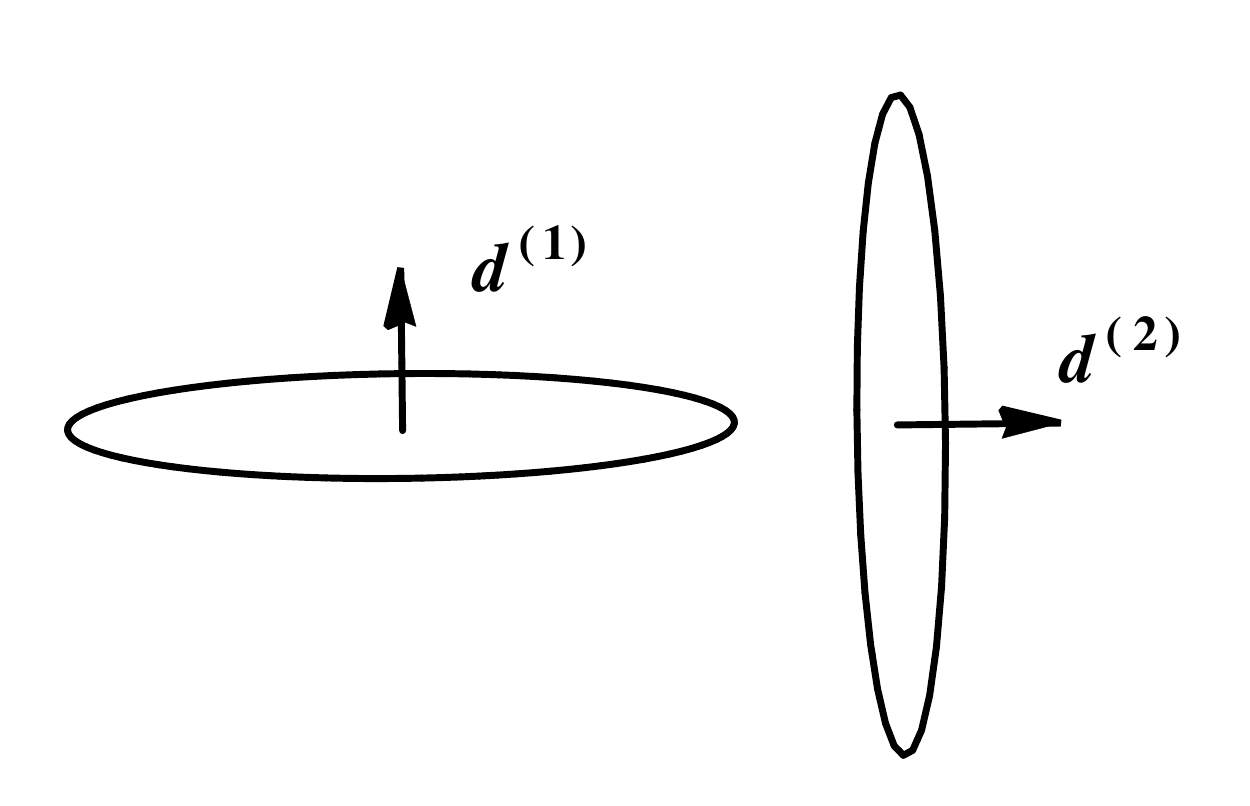}}
\end{minipage}}
\end{tabular}
\end{center}
\caption{The effective potential for three different, fixed configurations of the directors and the connecting vector. As a comparison we also plot the pair potential in the isotropic effective model. (a) $N=20$; (b) $N = 50$; (c) $N = 100$. The effective potentials are shown only for $r$ values for which we have relatively good statistics. We also show a sketch of the $||$, $--$ and $|-$ configurations.}
\label{vFixedPlots}
\end{figure}

The relative orientation between the vectors ${\bf r}$, $\mathbf{d}^{(1)}$ and $\mathbf{d}^{(2)}$ is of course not completely determined by the scalar product $\mathbf{d}^{(1)}\cdot \mathbf{d}^{(2)}$; the function ${\rm G}(r, \mathbf{d}^{(1)}\cdot \mathbf{d}^{(2)})$ contains less information than the full correlation function ${\rm g}(r,\cos \theta_1, \cos \theta_2, \varphi)$. In particular, when the directors are parallel, i.e., $\mathbf{d}^{(1)}\cdot \mathbf{d}^{(2)}=1$, the angle between the connection vector $\mathbf{r}$ and the directors $\mathbf{d}^{(1)}$ and $\mathbf{d}^{(2)}$ is still arbitrary. We denote a configuration with $\mathbf{d}^{(1)} \parallel \mathbf{d}^{(2)} \parallel \mathbf{r}$ as $--$ and a configuration with $\mathbf{d}^{(1)} \parallel \mathbf{d}^{(2)} \perp \mathbf{r}$ as $||$. From the reduced pair correlation function ${\rm G}(r, \mathbf{d}^{(1)}\cdot \mathbf{d}^{(2)})$ alone, we cannot say which of these two configurations is more probable, as the scalar product $\mathbf{d}^{(1)}\cdot \mathbf{d}^{(2)}$ is identical to $1$ in both cases. Using the full anisotropic pair correlation function ${\rm g}(r,\cos \theta_1, \cos \theta_2, \varphi)$ we can compare the corresponding effective potentials:
\begin{align}
\beta V_{--}(r)= -\ln\left[{\rm g}(r,\cos \theta_1 = 1, \cos \theta_2 = 1, \varphi)\right]\nonumber\\
\beta V_{||}(r)= -\ln\left[{\rm g}(r,\cos \theta_1 = 0, \cos \theta_2 = 0, \varphi=0)\right].
\end{align} 
For the $--$ case the value of the $\varphi$ coordinate is immaterial. However, due to the finite bin size of the grid on which we have calculated ${\rm g}(r,\cos \theta_1, \cos \theta_2, \varphi)$ the choice of $\varphi$ makes a small difference, even for $V_{--}(r)$. We compute $V_{--}(r)$ from the average of ${\rm g}(r,\cos \theta_1 = 1, \cos \theta_2 = 1, \varphi)$ in $\varphi$.

In figure \ref{vFixedPlots} we see that $V_{||}(r)$ increases significantly when $r$ approaches ${\rm D_{g0}}$, while $V_{--}$ stays close to 0 until much smaller distances $r$. We can understand this results if we imagine the rings as discs with diameter ${\rm D_{g0}}$. In the $||$ configuration, the rings lie in the same plane and will therefore start to overlap as soon as $r\leq {\rm D_{g0}}$. Since the rings are not perfect circles and their shape fluctuates, they can feel each other also for distances $r$ which are slightly larger than ${\rm D_{g0}}$. In the $--$ configuration two discs overlap only if the distance between their centers of mass is smaller than their thickness. These results tell us that the peak in the reduced pair correlation function ${\rm g}(r, \mathbf{d}^{(1)}\cdot \mathbf{d}^{(2)})$ for $r\approx 0.25 {\rm D_{g0}}$ and $\mathbf{d}^{(1)}\cdot \mathbf{d}^{(2)}\approx 1$ is mostly due to $--$ like configurations. However, as soon as the rings can overlap in $--$ type configurations the effective potential increases very fast for smaller $r$ and we come to a regime where other configurations of the directors are more favorable. As a comparison we also consider a configuration with $\mathbf{d}^{(1)} \perp \mathbf{d}^{(2)} \parallel \mathbf{r}$, which we denote by $|-$. The corresponding effective potential is given by
\begin{align}
\beta V_{|-}(r)= -\ln\left[{\rm g}(r,\cos \theta_1 = 0, \cos \theta_2 = 1, \varphi)\right].
\end{align} 
As in the $--$ case, the value of the $\varphi$ coordinate is irrelevant for calculating $V_{|-}(r)$, which we compute from the arithmetic mean of ${\rm g}(r,\cos \theta_1 = 0, \cos \theta_2 = 1, \varphi)$ in $\varphi$.

For small distances $r$ one ring interpenetrates the other in microscopic configurations of type $|-$. While $V_{|-}(r)$ starts to increase at larger $r$ values than $V_{--}(r)$, the increase is slower and converges to a constant for $r \rightarrow 0$. This is intuitive to understand since it requires only a finite amount of bending energy to deform two rings such that one can fit into the other. The required bending energy is smaller if the rings are larger. The $|-$ becomes dominant over the $--$ configuration at an $r$ value below the threshold $r\approx 0.25 {\rm D_{g0}}$. This is also the $r$ value at which we find the transition in the reduced pair correlation function ${\rm g}(r, \mathbf{d}^{(1)}\cdot \mathbf{d}^{(2)})$ between a regime where configurations with parallel directors, as in $--$, are preferred, to a regime where they are suppressed and other configurations like $|-$ become dominant.

\begin{figure}[htp]
\begin{center}

\begin{subfigure}[t]{8.5cm}
\includegraphics[width=8.5cm]{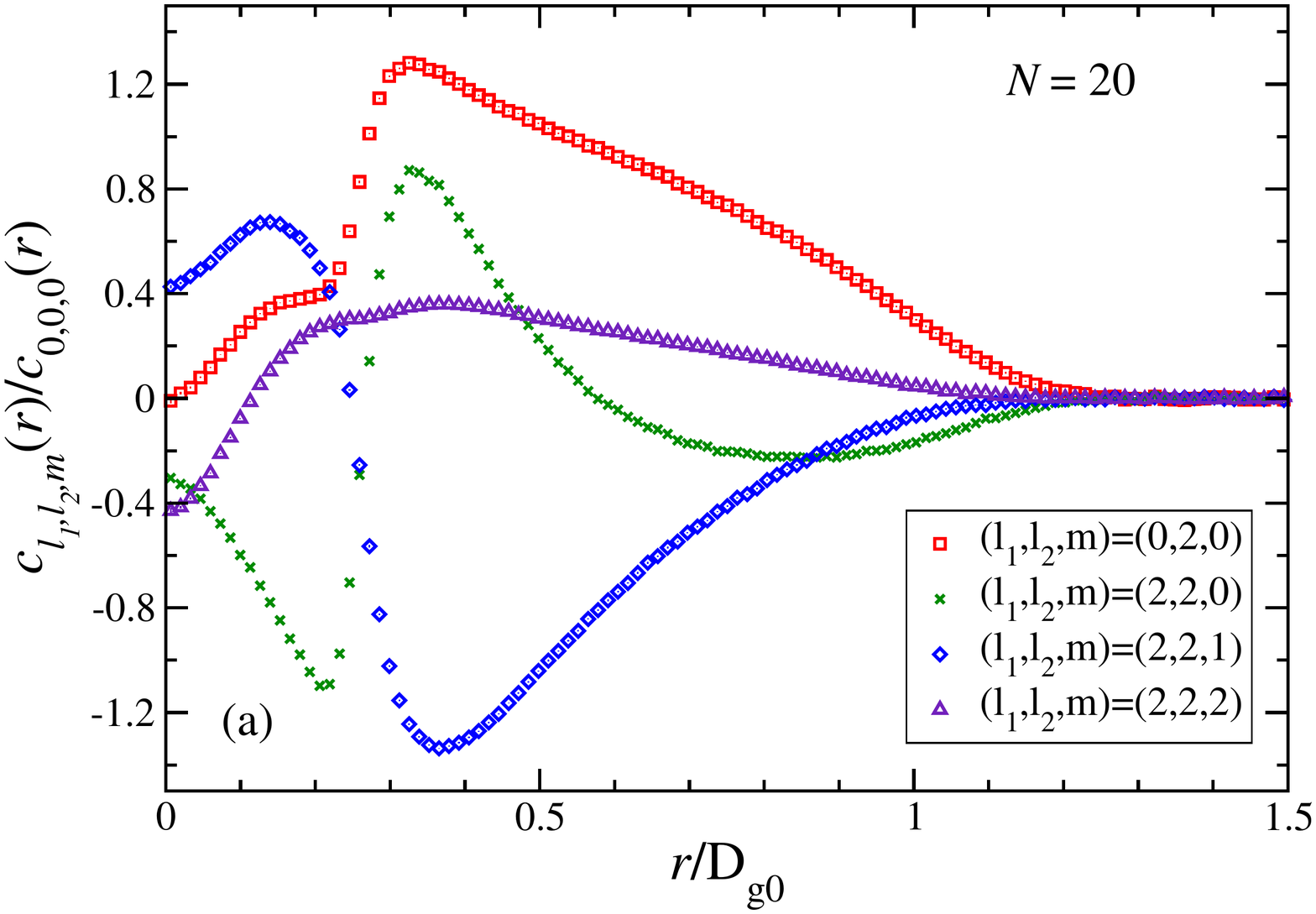}
\end{subfigure}
\begin{subfigure}[t]{8.5cm}
\includegraphics[width=8.5cm]{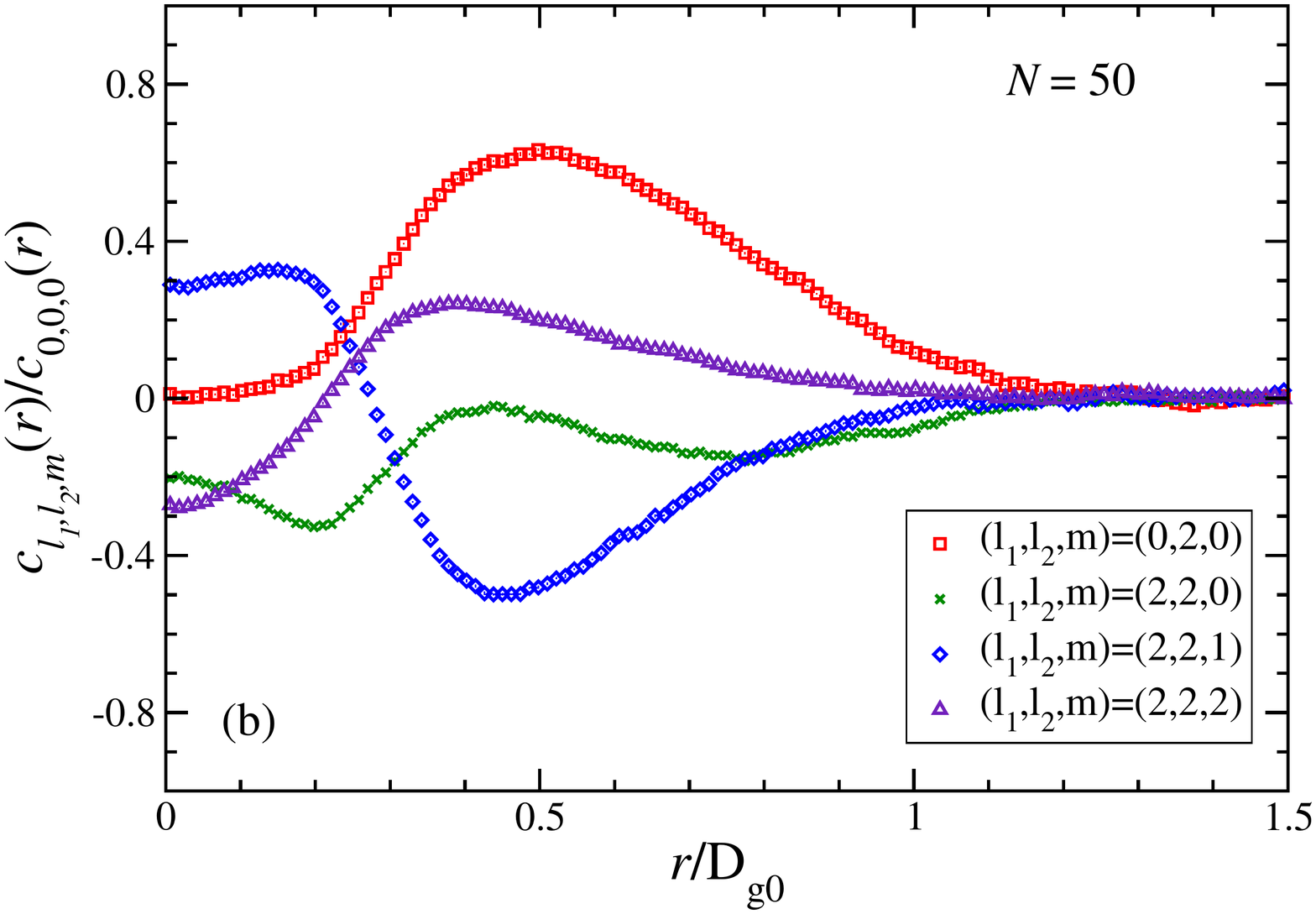} 
\end{subfigure}
\vspace{0.3cm}
\begin{subfigure}[t]{8.5cm}
\includegraphics[width=8.5cm]{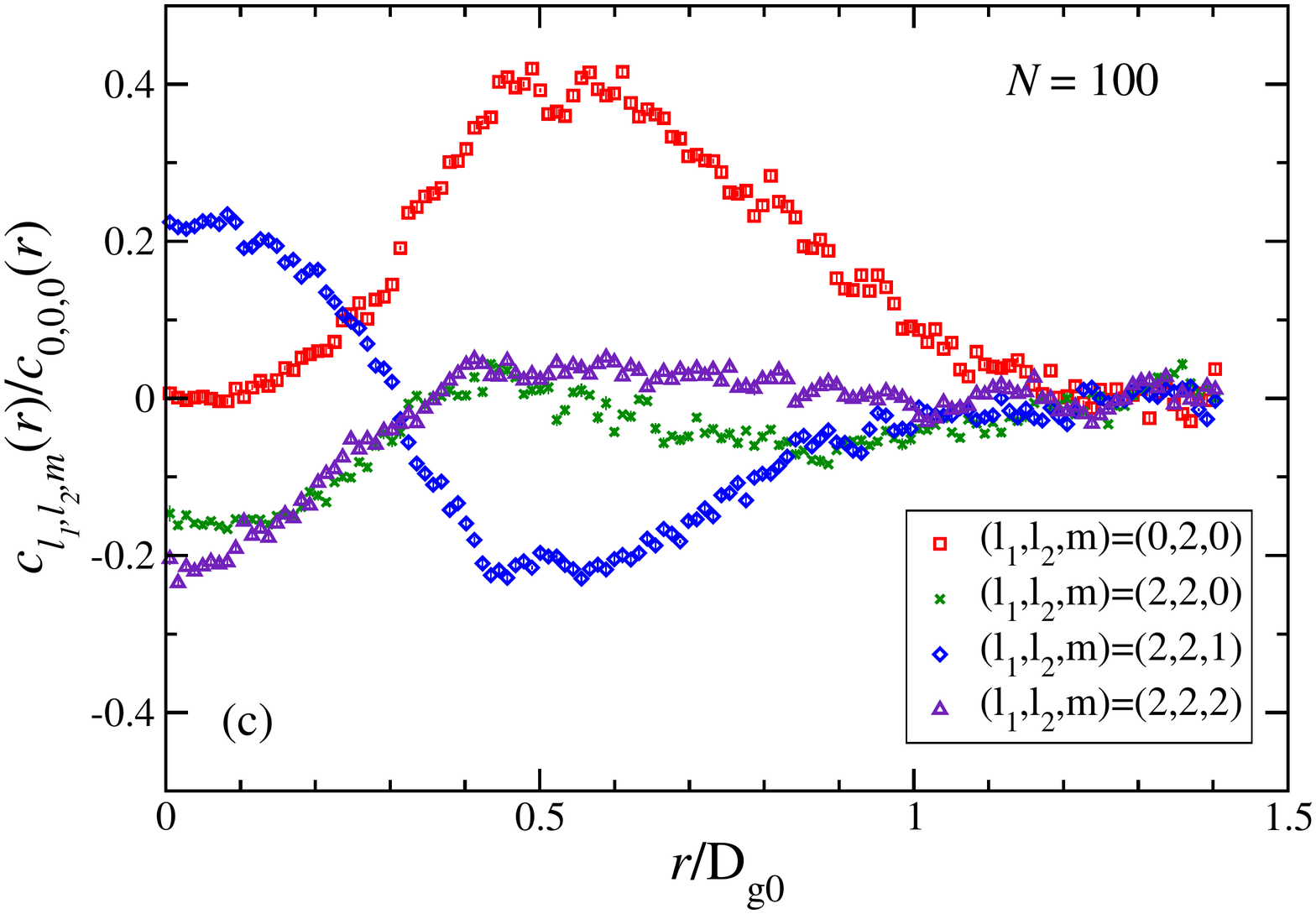} 
\end{subfigure}

\end{center}
\caption{The first coefficients in the expansion of ${\rm g}(r,\cos \theta_1, \cos \theta_2, \varphi)$ divided by the coefficent $c_{0,0,0}(r)$.
(a) $N = 20$; (b) $N = 50$; (c) $N = 100$.}
\label{angleCoeffPlots}
\end{figure}

In the Appendix we explain how one can expand the angular part of ${\rm g}(r,\cos \theta_1, \cos \theta_2, \varphi)$ into a series of suitably chosen basis functions ${\rm f}_{l_1,l_2,m}\left(\cos\theta_1,\cos\theta_2,\varphi\right)$. The expression for this expansion is given in Eq. (\ref{eq:expansionFormula}) and the corresponding coefficients ${\rm c}_{l_1,l_2,m}(r)$ can be determined by calculating particular ensemble averages as shown in Eq. (\ref{eq:cByAverage}). We plot these coefficients ${\rm c}_{l_1,l_2,m}(r)$ for $l_1,l_2\leq 2$ in figure \ref{angleCoeffPlots}. From the fast change of ${\rm c}_{l_1,l_2,m}(r)$ for $r\approx 0.25 {\rm D_{g0}}$ one can once more see the transition between two regimes for $r$ in which the distribution of the directors of the ring polymers is very different. We can again see that this transition is smoother for larger rings. The magnitude of coefficients ${\rm c}_{l_1,l_2,m}(r)/{\rm c}_{0,0,0}(r)$ with $(l_1,l_2)\neq (0,0)$ tells us about the significance of the corresponding anisotropy in ${\rm g}(r,\cos \theta_1, \cos \theta_2, \varphi)$. Anisotropy is more important for smaller rings and becomes more pronounced after the transition at $r \approx 0.25 {\rm D_{g0}}$, where the rings prefer parallel configurations.

\section{Monte Carlo Simulations of the Anisotropic Effective Model}
\label{manybody:sec}

\begin{figure}[htp]
\begin{center}

\begin{subfigure}[t]{8.5cm}
\includegraphics[width=8.5cm]{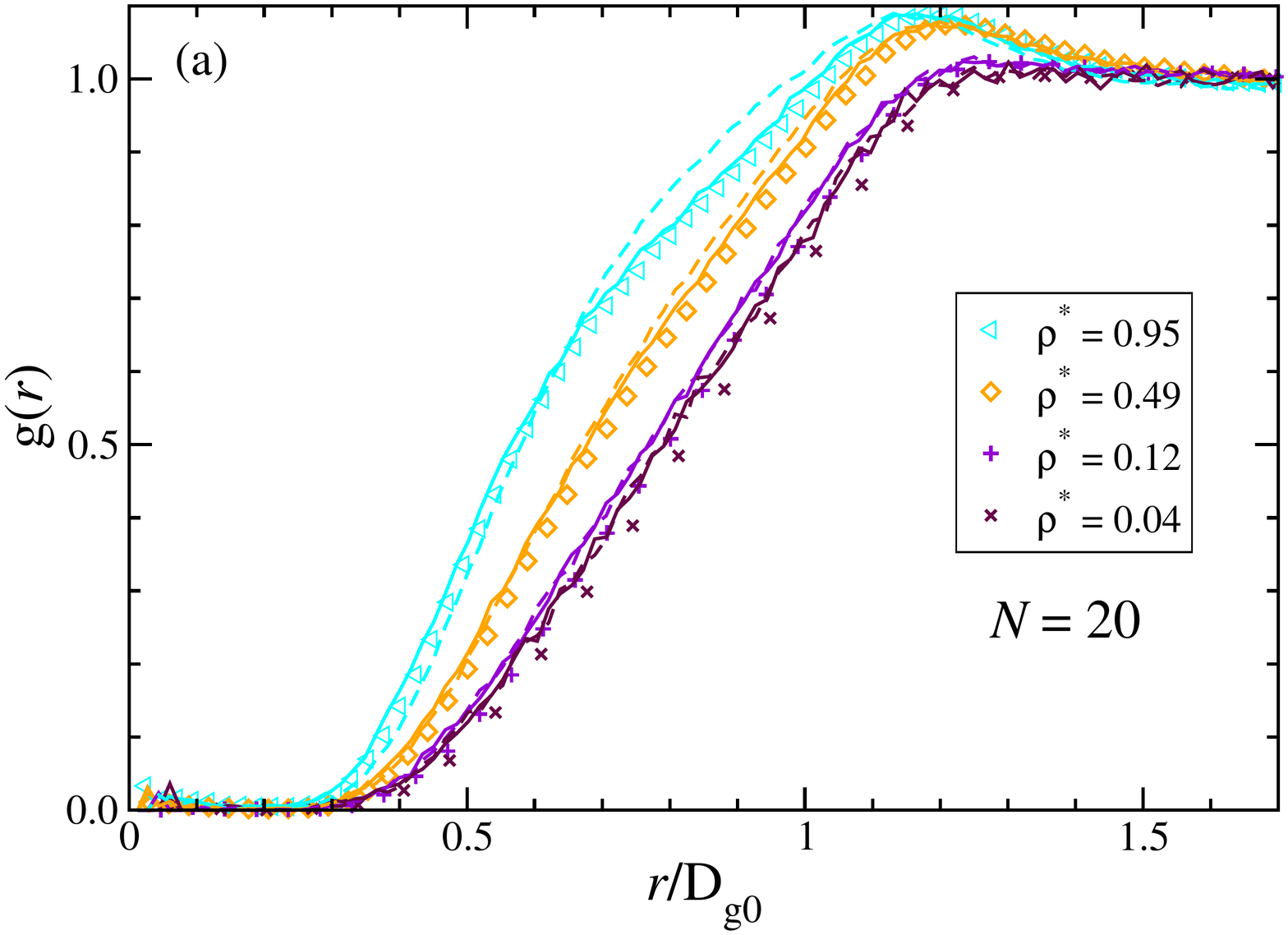}
\end{subfigure}
\begin{subfigure}[t]{8.5cm}
\includegraphics[width=8.5cm]{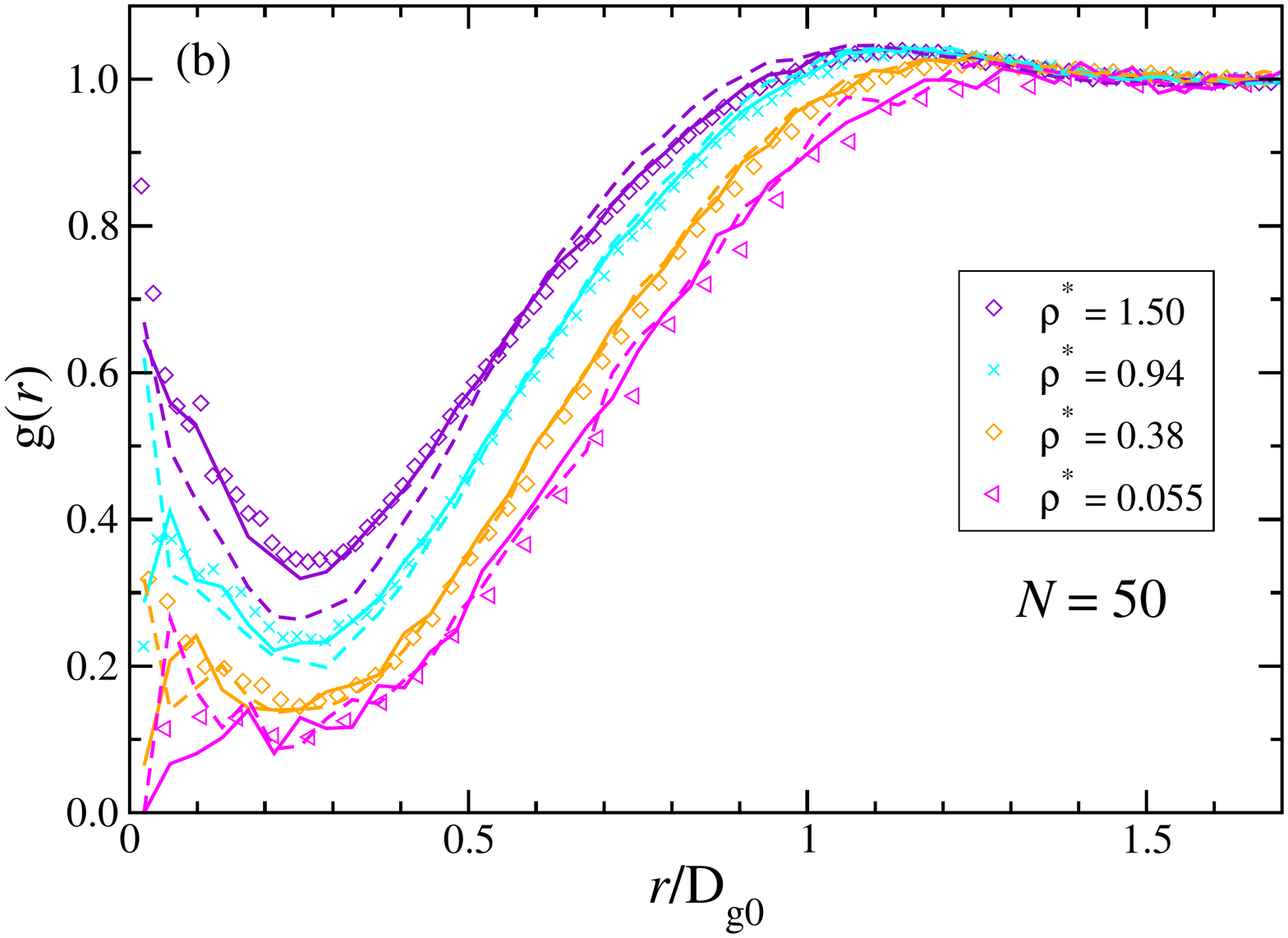} 
\end{subfigure}
\vspace{0.3cm}
\begin{subfigure}[t]{8.5cm}
\includegraphics[width=8.5cm]{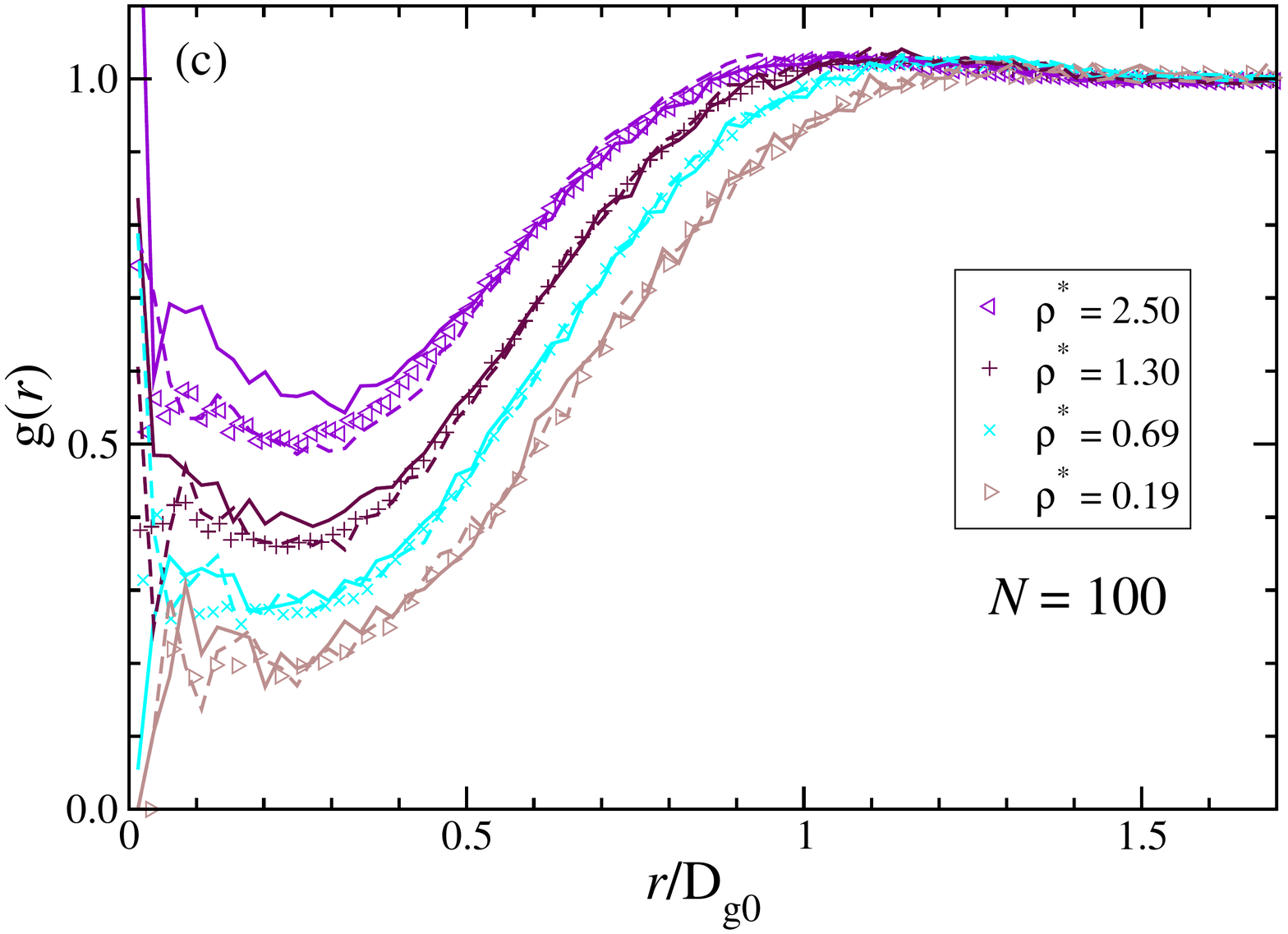} 
\end{subfigure}

\end{center}
\caption{The pair correlation function ${\rm g}(r)$ at low reduced densities $\rho^*$ for a simulation of many ring polymers in the full monomer-resolved simulation (symbols), the anisotropic effective model (solid line) and the isotropic effective model (dashed line).}
\label{gREffSimCheck}
\end{figure}

We carried out Monte Carlo simulations of systems of effective particles described by the anisotropic effective model for different ring sizes $N$ and various densities. We define the reduced density in our simulation as $\rho^*\equiv n {\rm D_{g0}^3}/ L^3$, where $n$ is the number of rings in the sample. In order to assess the quality of the anisotropic effective model, we compare our results to results of full monomer-resolved simulations from ref. \cite{bernabei2013fluids} and the results of simulations using the isotropic effective model. As we can see in figure \ref{gREffSimCheck} for all choices of the number of monomers $N$ the effective models are in good agreement with the full monomer-resolved simulations at low densities $\rho^*$. This is an important consistency check for the effective models, in which the interactions have been chosen such that the distribution of the effective degrees of freedom agrees with their distribution in the full monomer-resolved simulations, in the limit of small densities. 

\begin{figure}[htp]
\begin{center}
\includegraphics[width=12.0cm]{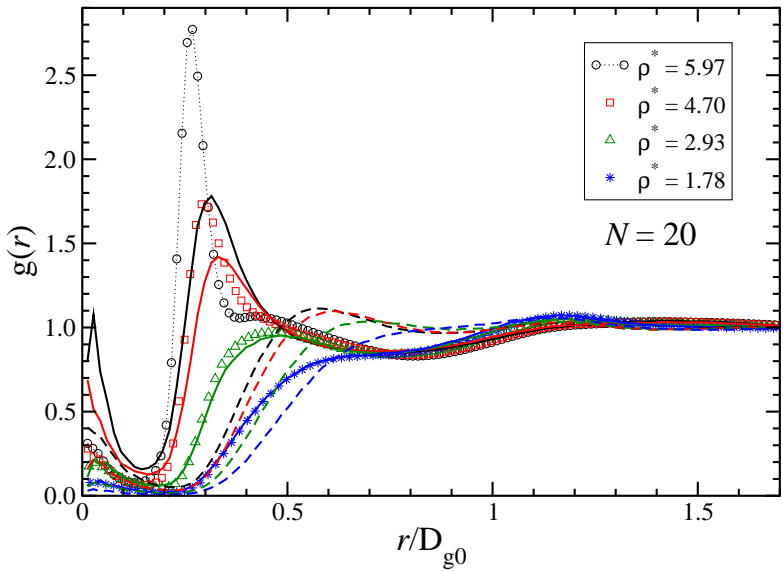} 
\end{center}
\caption{The pair correlation function ${\rm g}(r)$, at high densities, for a simulation of many ring polymers with $N=20$ monomers in the full monomer-resolved simulation (symbols), the anisotropic effective model (solid line) and the isotropic effective model (dashed line).}
\label{gREffSimN20Agree}
\end{figure}

In figure \ref{gREffSimN20Agree}, we present results for the smallest rings with $N=20$ monomers at higher densities.
There is a dramatic improvement of the accuracy as one compares the isotropic with the anisotropic model.
While the former fails for $\rho^* > 2$ the anisotropic effective model works up to $\rho^* \cong 5$ and even gives a semi-quantitatively correct description of the system at $\rho^*= 5.97$. At the highest densities, we see the development of a peak in ${\rm g}(r)$ at $r \cong 0.3 {\rm D_{g0}}$. This peak in the pair correlation function is associated with the emergence of stacks of parallel rings and its position describes the typical distance of rings in these stacks\cite{bernabei2013fluids}. Interestingly, the isotropic effective potential has a \textit{maximum} for $r \approx 0.25 {\rm D_{g0}}$ which is close to the typical distance of the rings in the stacks and one could wonder why the rings prefer to align at a distance which seems to have a very high free energy penalty according to the isotropic effective potential. The answer to this apparent paradox lies in the strongly peaked nature of the anisotropic effective interaction, which we could observe in figures \ref{gD1DotD2Plots}, \ref{vFixedPlots} and \ref{angleCoeffPlots}. While the average configuration of the angular degrees of freedom at distances $r\approx 0.3 {\rm D_{g0}}$ has a high free energy penalty, a certain class of configurations, where the directors of the rings are almost parallel, is much more favorable. Obviously, stacking can not be observed in the isotropic effective model, where particles possess no directional degrees of freedom.

\begin{figure}[htp]
\begin{center}
\includegraphics[width=12.0cm]{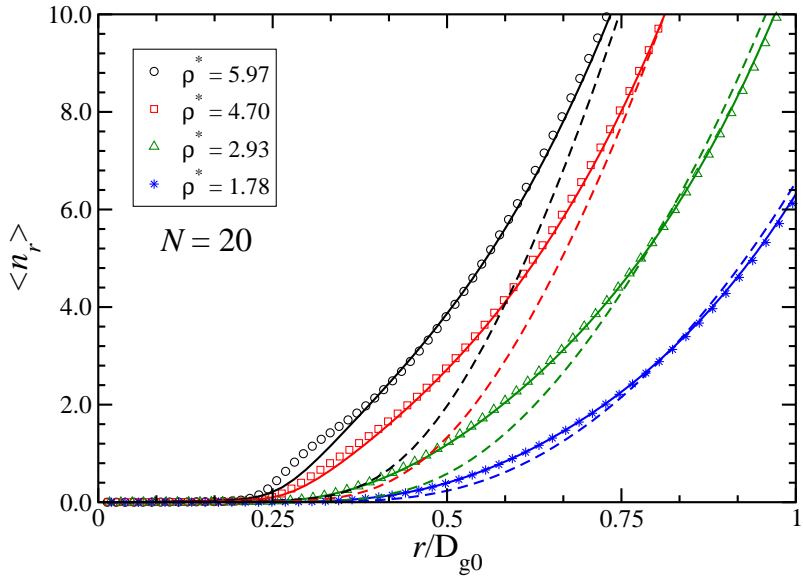} 
\end{center}
\caption{$\langle n_r\rangle$ in a simulation of many ring polymers with $N=20$ monomers in the full monomer-resolved simulation (symbols), the anisotropic effective model (solid line) and the isotropic effective model (dashed line).}
\label{gREffSimN20CumAgree}
\end{figure}

As a further characteristic of the short-range coordination of the rings, we consider the average number $\langle n_r\rangle$ of neighbors within a distance $r$ from 
the center of mass of a randomly chosen ring. This is expressed as
\begin{eqnarray}
\langle n_r\rangle = 4\pi\rho\int_0^r{\rm d}x\, x^2 {\rm g}(x).
\label{nofr:eq}
\end{eqnarray}
For the rings with $N=20$ monomers we present results for $\langle n_r\rangle$ in figure \ref{gREffSimN20CumAgree}. Once more, 
the good agreement between the full monomer-resolved and the anisotropic effective model, even at the highest densities investigated, is confirmed: small differences appear only for $0.25 {\rm D_{g0}}\leq r \leq 0.4 {\rm D_{g0}}$. Evidently, $\langle n_r\rangle$ does not contain more information than the ${\rm g}(r)$ plot in figure \ref{gREffSimN20Agree}, but it nevertheless clarifies the meaning of the disagreement between the ${\rm g}(r)$ curves in the monomer-resolved and the anisotropic effective model. At the highest densities in the full simulation, the centers of mass move a bit closer to each other than they do in the anisotropic effective simulation. This manifests itself as a shift of the peaks in the ${\rm g}(r)$ curves. The difference in the height of the peaks is partly a consequence of the shift, since a peak in ${\rm g}(r)$ has to be higher at smaller distances if it amounts to the same amount of average neighbours as a peak at a larger distance $r$. The fact that the $\langle n_r\rangle$ curves for the anisotropic effective and the full simulation in figure \ref{gREffSimN20CumAgree} agree for $r \geq 0.4 {\rm D_{g0}}$ shows us that the peaks in the ${\rm g}(r)$ curves indeed correspond to the same amount of average nearby particles that are simply accumulated at slightly different distances.

\begin{figure}[htp]
\begin{center}
\begin{subfigure}[t]{8.5cm}
\includegraphics[width=8.5cm]{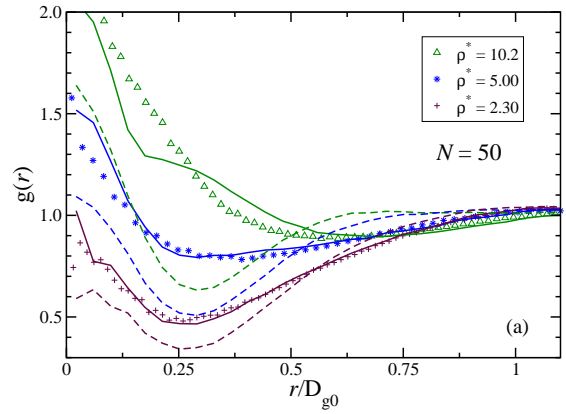}
\end{subfigure}
\begin{subfigure}[t]{8.5cm}
\includegraphics[width=8.5cm]{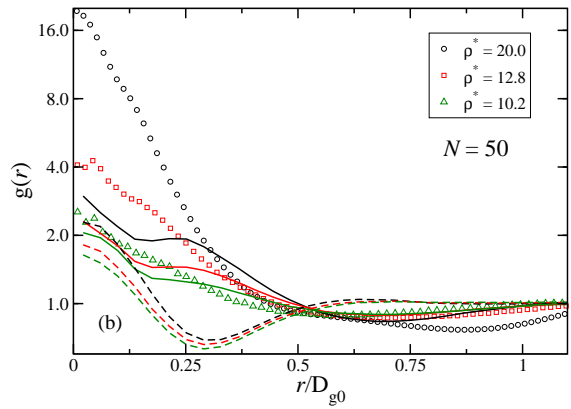} 
\end{subfigure}
\end{center}
\caption{The pair correlation function ${\rm g}(r)$ for a simulation of many ring polymers with $N=50$ monomers in the full monomer-resolved simulation (symbols), the anisotropic effective model (solid line) and the isotropic effective model (dashed line). The two plots show different reduced density $\rho^*$ ranges.}
\label{gREffSimN50AgreeHigh}
\end{figure}

We proceed now with the longer rings, $N = 50$.
As can be seen by the pair correlation curves in figure \ref{gREffSimN50AgreeHigh}(a), also in this case the inclusion of anisotropy improves the agreement with the monomer-resolved simulations significantly for densities $\rho^*$ from $2.3$ to $10.2$. In figure \ref{gREffSimN50AgreeHigh}(b) we present ${\rm g}(r)$ for higher $\rho^*$. In the full monomer-resolved simulations we can see a peak emerging in the pair correlation function ${\rm g}(r)$ at $r=0$ on increasing the density. As described in ref. \cite{bernabei2013fluids,Slimani2014} the monomer-resolved system forms stacks of quasi-parallel oblate rings that are fully penetrated by bundles of elongated rings. In this phase, the deformation of the penetrating rings is particularly strong. The effective description, on the other hand, breaks down if the internal configurations of the rings in the monomer-resolved system differ significantly to the internal configurations in the system with only 2 ring polymers. Therefore, the anisotropic effective model should not be expected to be a quantitative description at the high densities in which this phase is formed.
Agreement with the monomer-resolved model here is less satisfactory but the improvement over the isotropic model is still spectacular.

\begin{figure}[htp]
\begin{center}
\begin{subfigure}[t]{8.5cm}
\includegraphics[width=8.5cm]{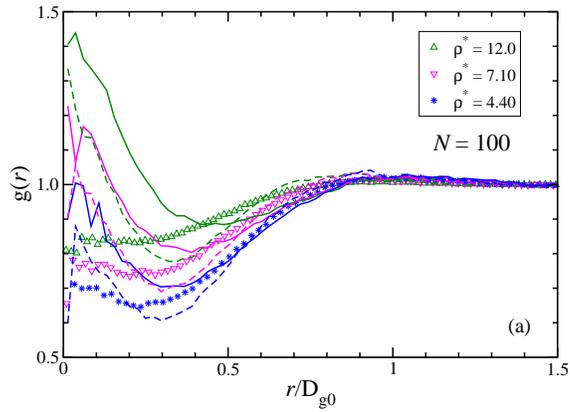}
\label{gREffSimN100High1}
\end{subfigure}
\begin{subfigure}[t]{8.5cm}
\includegraphics[width=8.5cm]{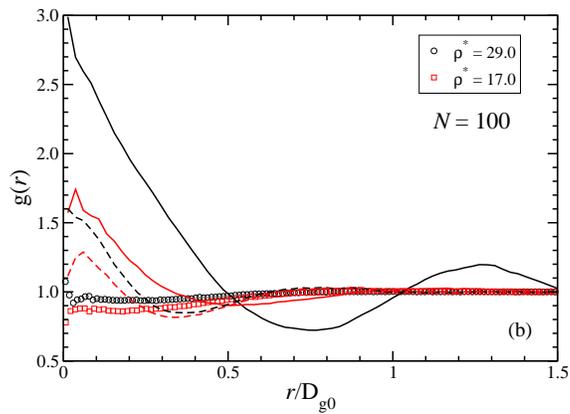} 
\label{gREffSimN100High2}
\end{subfigure}
\end{center}
\caption{The pair correlation function ${\rm g}(r)$ for a simulation of many ring polymers with $N=100$ monomers in the full monomer-resolved simulation (symbols), the anisotropic effective model (solid line) and the isotropic effective model (dashed line). The two plots show different reduced density $\rho^*$ ranges.}
\label{gREffSimN100High}
\end{figure}

For the rings with $N=100$ we find that anisotropy does not play a key role any more, at least not for the full model at the investigated densities. This had to be expected, as we could already see in figure \ref{gD1DotD2Plots} and \ref{angleCoeffPlots} that anisotropy is less pronounced for larger ring sizes. As we saw in figure \ref{gREffSimCheck}(c) both the isotropic and the anisotropic model give good results for ${\rm g}(r)$ up to $\rho^*\approx 2.5$. In figure \ref{gREffSimN100High}, we see that for higher densities the inclusion of anisotropy does not yield results that are in better agreement with the full monomer-resolved simulations. The results in the isotropic effective model even seem to be in better agreement with the full model, which is attributed to multi-particle interactions that can change the configurations of the large and therefore more deformable rings significantly. The already small correlation between the directors, which is present in the dilute case, might therefore be even smaller at high densities. In the anisotropic effective model, we then overestimate the angular correlations between the directors and arrive at results that can be slightly worse than those of the isotropic model. Interestingly at $\rho^*=20.0$, which is the highest density investigated, the anisotropic model appears to crystallize. At this density we see the emergence of columns that are closed over the periodic boundary conditions and organize in a hexagonal 2D lattice structure.

\begin{figure}[htp]
\begin{center}

\begin{subfigure}[t]{8.5cm}
\includegraphics[width=8.5cm]{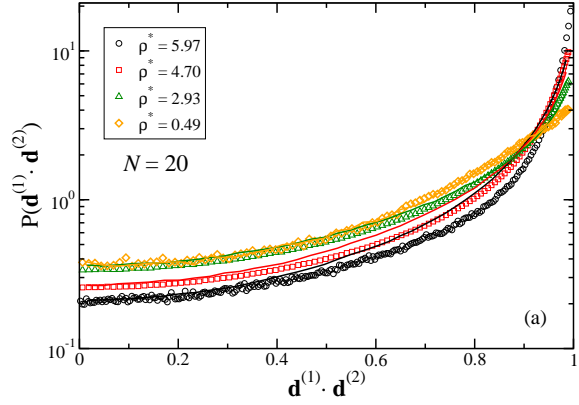}
\label{pCosThetaEffSimN20}
\end{subfigure}
\begin{subfigure}[t]{8.5cm}
\includegraphics[width=8.5cm]{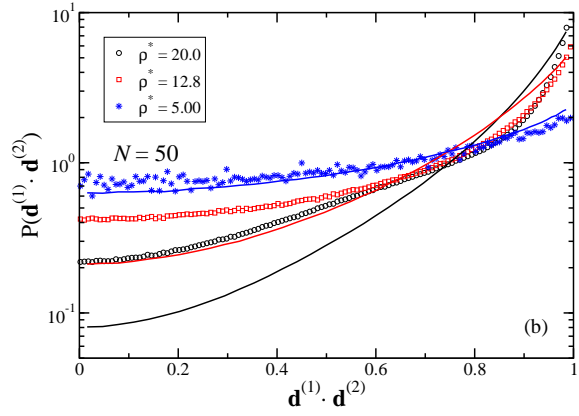} 
\label{pCosThetaEffSimN50}
\end{subfigure}
\vspace{0.3cm}
\begin{subfigure}[t]{8.5cm}
\includegraphics[width=8.5cm]{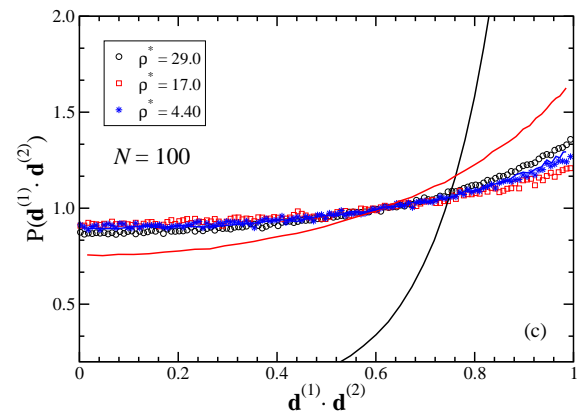} 
\label{pCosThetaEffSimN100}
\end{subfigure}
\end{center}
\caption{${\rm P}(\mathbf{d}^{(1)} \cdot \mathbf{d}^{(2)})$ is the probability density to find the scalar product $\mathbf{d}^{(1)} \cdot \mathbf{d}^{(2)}$ between the directors of two close by rings ($r< 0.6 {\rm D_{g0}}$). Here we show ${\rm P}(\mathbf{d}^{(1)} \cdot \mathbf{d}^{(2)})$  for a simulation of many ring polymers in the full monomer-resolved simulation (symbols) and the anisotropic effective model (solid line). (a) $N = 20$; (b) $N = 50$; (c) $N = 100$.}
\label{pThetaEffSim}
\end{figure}

Finally, let us focus exclusively on orientational correlations. 
We define ${\rm P}(\mathbf{d}^{(1)} \cdot \mathbf{d}^{(2)})$ as the probability density distribution for the scalar products between the directors 
$\mathbf{d}^{(1)}$ and $\mathbf{d}^{(2)}$
of two ring polymers which are a distance $r<0.6 {\rm D_{g0}}$ away from each other. In figure \ref{pThetaEffSim}, we present results for ${\rm P}(\mathbf{d}^{(1)} \cdot \mathbf{d}^{(2)})$ for simulations in the monomer resolved and in the anisotropic effective model. 
If the directors were uncorrelated ${\rm P}(\mathbf{d}^{(1)} \cdot \mathbf{d}^{(2)})$ would be equal to $1$.
For low densities $\rho^*$ we obtain good agreement for all ring sizes investigated. Since we only look at the directional correlation of close by ring polymers, ${\rm P}(\mathbf{d}^{(1)} \cdot \mathbf{d}^{(2)})$ can show strong anisotropic features even for $\rho^* \rightarrow 0$. As expected the anisotropy in ${\rm P}(\mathbf{d}^{(1)} \cdot \mathbf{d}^{(2)})$ is  stronger for smaller rings. When the density is increased, the distribution always shifts towards parallel configurations in the effective model. This happens because less volume per ring is available for higher $\rho^*$ and by aligning parallel the rings occupy less space. Typically one observes the same trend in the monomer resolved simulation, only for the $N=100$ rings we find more parallel rings for $\rho^*=2.5$ than for $\rho^*=17.0$. In contrast to the effective model the rings in the monomer-resolved simulation can deform and their interaction with other rings can therefore be more isotropic at higher densities $\rho^*$. This explains why for the large rings with $N=100$ monomers, which deform more easily than the smaller rings, the correlation between the directors is much weaker than in the effective model and can even decrease with density. For $N=50$ one can see that the number of orthogonal rings in the monomer-resolved model at high densities is significantly larger than in the effective simulation. As described in ref. \cite{bernabei2013fluids,Slimani2014} for $N=50$ and $\rho^*\geq 12.8$ one observes that oblate rings are interpenetrated by elongated prolate rings. Since the directors of the oblate and the interpenetrating prolate rings can be orthogonal to each other, one observes perpendicular directors for $N=50$ even at the highest densities investigated. In the anisotropic effective model on the other hand, this interpenetration is disfavoured and we observe almost no orthogonal close-by rings at $\rho^*=20$ for $N=50$.

\begin{figure}[htp]
\begin{center}

\begin{subfigure}[t]{8.5cm}
\includegraphics[width=8.5cm]{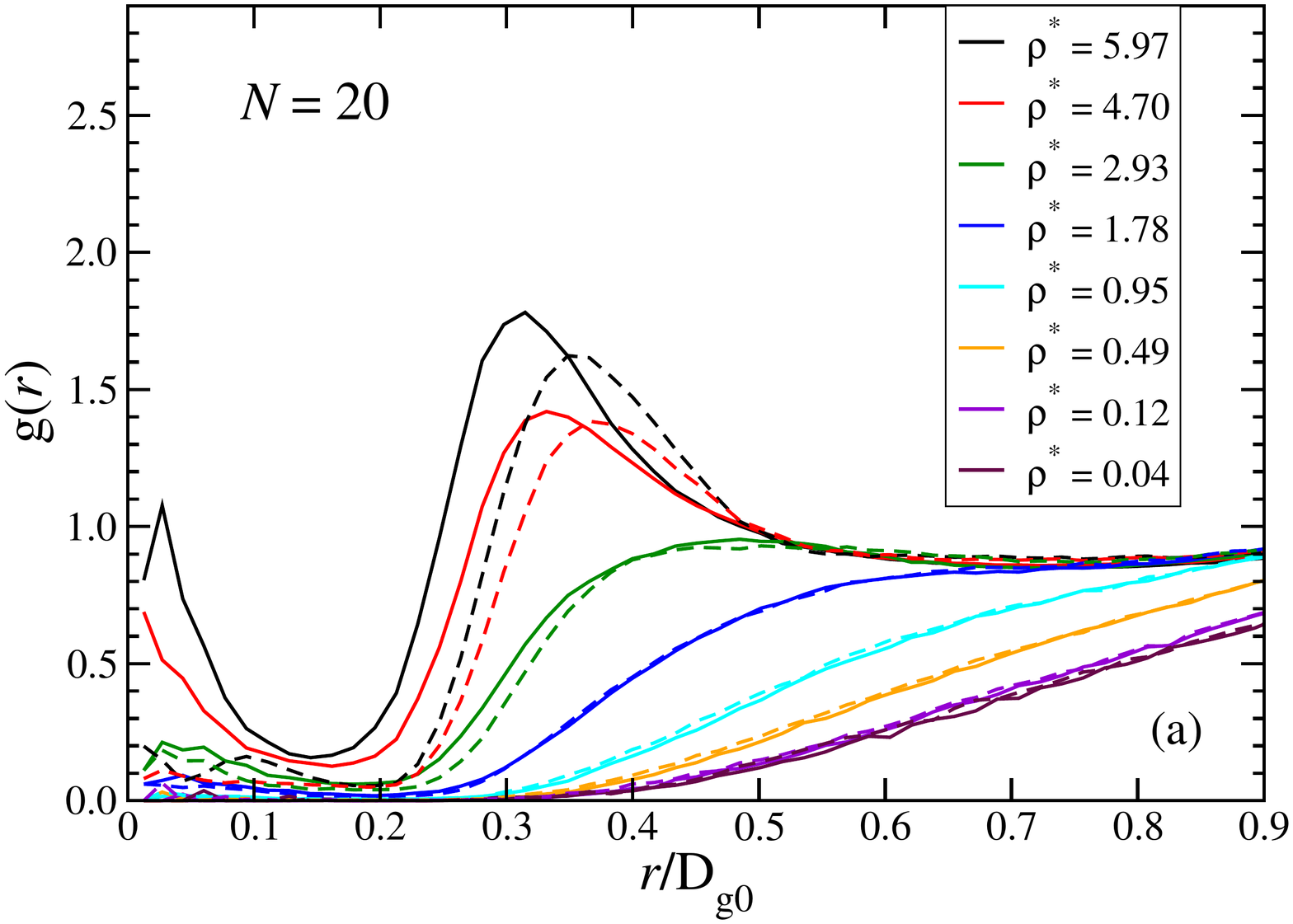}
\end{subfigure}
\begin{subfigure}[t]{8.5cm}
\includegraphics[width=8.5cm]{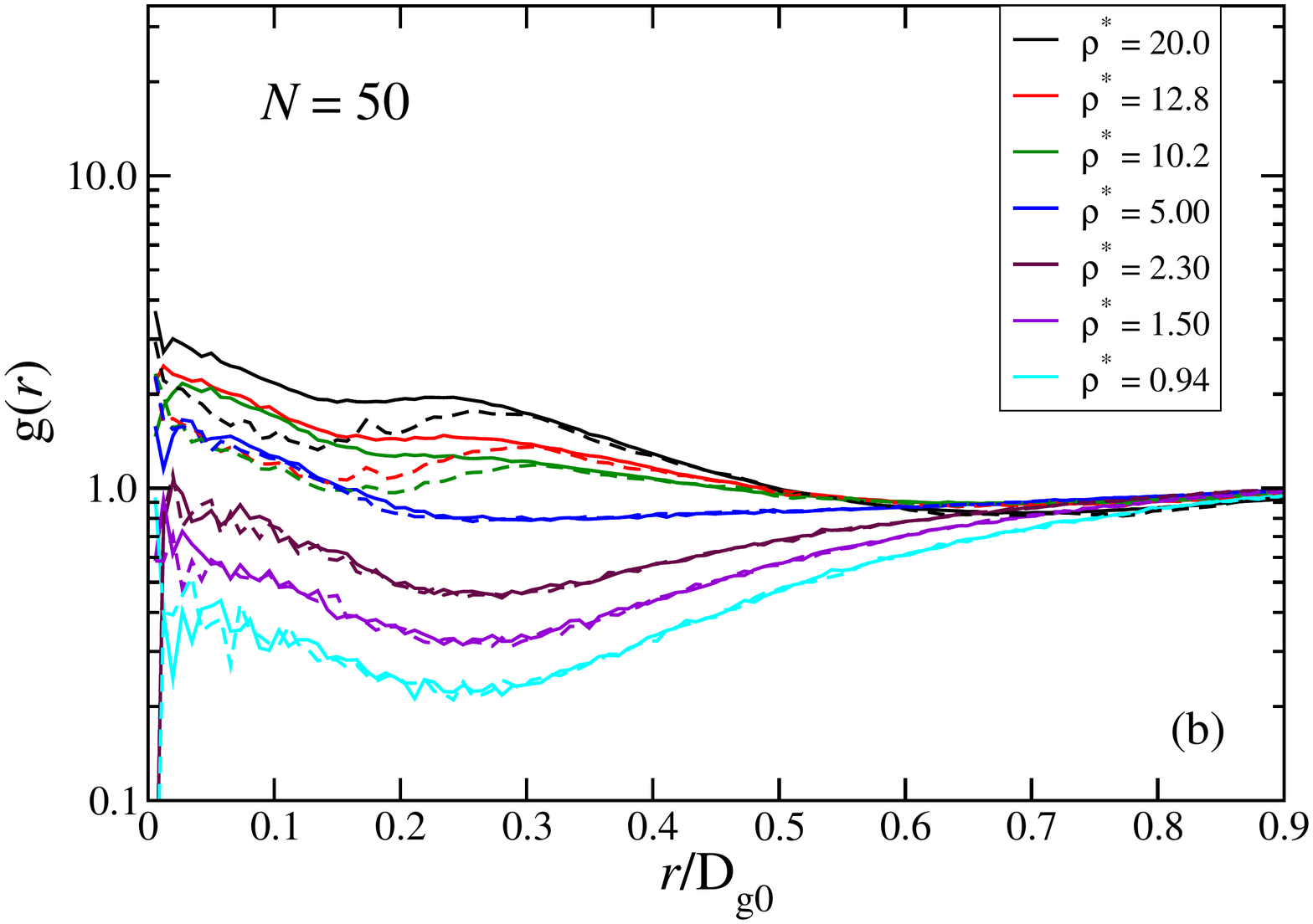} 
\end{subfigure}

\end{center}
\caption{The pair correlation function ${\rm g}(r)$ for a simulation of many ring polymers in the anisotropic effective model. For the dashed line we expanded the pair-correlation function ${\rm g}$ before computing the associated effective pair-potential. For the expansion we took the 14 coefficients for which $l_1,l_2\leq 4$ into account. The solid line shows results of a simulation with the unexpanded effective pair interaction. (a) $N = 20$; (b) $N = 50$.}
\label{gREffExpCmpr}
\end{figure}

\section{Truncation of the Expansion of the Anisotropic Potential}
\label{truncate:sec}

Instead of working with a fully tabulated effective potential on a four-dimensional grid, it can be advantageous to use the analytical expansion on basis
functions presented in the Appendix. Such expansions are truncated after some term, and here we shortly discuss the quality of such truncations for the problem at hand.
To test the quality of the expansion of ${\rm g}\left(r,\cos\theta_1,\cos\theta_2,\varphi\right)$ we also carried out Monte Carlo Simulations, where we used the effective potential
associated to the expanded correlation function
as the pair-interaction between our effective particles. We took the 14 coefficients ${\rm c}_{l_1,l_2,m}(r)$ for which $l_1,l_2\leq 4$ into account
and truncated the rest of the expansion. While ${\rm g}\left(r,\cos\theta_1,\cos\theta_2,\varphi\right)$ can never be negative, the truncated expanded version of ${\rm g}$ can accidentally become smaller than zero. Wherever this happens ${\rm g}=0$ and ${\rm V_{eff}}=\infty$ is used in the simulation. In figure \ref{gREffExpCmpr} the pair correlation function ${\rm g}(r)$ obtained in this simulation is shown in comparison with the ${\rm g}(r)$ function, which we computed previously employing the full anisotropic effective potential. For both $N=20$ and $N=50$ we obtain reasonable results with the truncated effective interaction, given by only 14 expansion coefficients. For the full effective interaction, which we store on a 4D grid, we save $16^3=4096$ entries for each value of $r$ (see section \ref{subsec:derivEffInt}). At intermediate densities the results obtained with the expanded effective interaction are a significant improvement with respect to the isotropic effective model. However, one has to be aware that for $N=20$ the coefficients of higher order modes can still be quite high, especially for $r$ between $0.2 {\rm D_{g0}}$ and $0.7 {\rm D_{g0}}$. In figure \ref{angleCoeffPlots} we see that the coefficient for the mode with $(l_1,l_2,m)=(0,2,0)$ can be larger than the coefficient of the isotropic expansion mode.
%The same is true for the mode with $(l_1,l_2,m)=(4,4,0)$.
The reason for the high contribution of higher order modes for $N=20$ is of course the strongly peaked nature of ${\rm g}\left(r,\cos\theta_1,\cos\theta_2,\varphi\right)$ for these small rings, which we can also observe in figure \ref{gD1DotD2Plots}. The convergence of ${\rm g}\left(r,\cos\theta_1,\cos\theta_2,\varphi\right)$ is poor for the $N=20$ rings due to the strong anisotropy of their effective interaction. However our results show that the expansion modes up to $l_1,l_2\leq 4$ already capture the main features of the effective interaction. For $N=50$ the degree of anisotropy is weaker and therefore the convergence of the expansion of ${\rm g}\left(r,\cos\theta_1,\cos\theta_2,\varphi\right)$ is better.
    
\section{Conclusions}
\label{sec:conclusions}

We have introduced a minimal anisotropic model to coarse-grain ring polymers with a finite bending rigidity as soft, penetrable disks. For the shortest ($N = 20$)
and the intermediate ($N = 50$) sized rings, this model represents a dramatic improvement over the isotropic coarse-graining, in which the relative
orientations between the rings are all integrated upon and a radially symmetric interaction results instead. The approach is capable of distinguishing
between the relative orientations at infinite dilution and it carries this distinction also to highly concentrated systems, where it reproduces well the salient
features of the structure as seen in the full, monomer-resolved simulations. Whereas this is valid more for $N = 20$ and $N = 50$, which have a contour length to persistence length ratio of $N/s_p\sim 2.7$ and $6.7$ respectively, some important features,
such as the penetration of elongated rings in columns formed by  oblate rings (found for $N=50$), are suppressed or even lost in the effective description, as genuine many-body
effects come into play. For the largest rings, $N = 100$, for which we obtain $N/s_p\sim13.3$, the contour length is much larger than the persistence length and they thus resemble more
flexible objects. In this case the anisotropic potential at high concentrations fails to describe the structural correlations. This indeed reflects the fact that such rings
undergo, at high concentrations, conformational changes (shrinking, interpenetration) that are quite distinct from the assumptions that go into the anisotropic,
soft disc-model, rendering it thereby very inaccurate. We therefore expect that our anisotropic model yields quantitative results over a broad density range, for systems of polymer rings with a contour to persistence length ratio of $N/s_p\lesssim 10$. 

Our work provides, thus, an accurate and efficient general scheme for the coarse-graining of semiflexible ring polymers, as it allows for a very dramatic
reduction of their degrees of freedom, while at the same time introducing a realistic class of systems for which anisotropic generalizations of the
ultrasoft, penetrable effective interactions are physically meaningful. Future work will focus on the investigations of the structural and phase behavior
of mixtures of stiff rings and of the dynamics of the structure formation in the same.

\begin{acknowledgement}
This work has been supported by the Austrian Science Fund (FWF), Grant 23400-N16.
\end{acknowledgement}

\appendix

\section{Calculation of the Anisotropic Effective Potential}
\label{sec:SelfConsistentHisto}
As discussed in section \ref{sec:AnisoEffModel}, we wish to sample ${\rm P}\left(r,\cos\theta_1,\cos\theta_2,\varphi\right)$ for a system of two ring polymers, in order to obtain a numerical expression for the anisotropic effective potential between them. We know that for large values of $r$, when the polymers can not interact with each other, ${\rm P}$ will correspond to the ideal case, ${\rm P_{id}}$. Therefore the interesting configurations for us are those values of $r$ that result in overlaps between the ellipsoids of gyration. We use a biasing potential between the centers of mass of the two rings to restrict $r$ to certain \textit{umbrella windows}:
\begin{eqnarray}
\label{eq:VBias}
{\rm V}_{\rm bias}^{(j)}(r)= \frac{k_j}{2} \left(r-r_j\right)^2.
\end{eqnarray}
With $r_j$, $k_j$ we can tune respectively the location and width of the window for $r$ in which configurations are sampled. We carry out simulations for a range of different $r_j$ and $k_j$ values and calculate histograms ${\rm P}^{(j)}_\text{bias}(Q)$ in the effective coordinates. Here, $Q$ stands for $(r,\cos \theta_1, \cos \theta_2, \varphi)$ as a collective variable and thus denotes a bin in the effective coordinates, whereas $j$ is the index of the respective biased simulation and thus determines $k_j$ and $r_j$. The binning in the 4D space is identical for all biased simulations. We use the \textit{Self-Consistent Histogram Method} by Ferrenberg and Swendsen\cite{ferrenberg1989optimized,frenkel2001understanding} to combine the different ${\rm P}^{(j)}_\text{bias}(Q)$, which results in an estimate ${\rm P}_{{\rm est},j}(Q)$ for the histogram ${\rm P}(Q)$ of the unbiased system.
The starting point of this method is that every simulation does in principle give an estimate for the histogram ${\rm P}(Q)$ of the unbiased simulation:
\begin{eqnarray}
{\rm P}_{{\rm est},j}(Q)= N^{(j)} \exp\left(\beta {\rm V}_{\rm bias}^{(j)}(Q)\right) {\rm P}^{(j)}_\text{bias}(Q).
\label{eq:pEst0IDef}
\end{eqnarray}
Here, ${\rm V}^{(j)}_{\rm bias}$ denotes the bias potential in the $j$-th simulation, given by (\ref{eq:VBias}) and $N^{(j)}$ is a normalization factor, which can be expressed as
\begin{eqnarray}
N^{(j)}= \sum_Q {\rm P}(Q) \exp\left(-\beta {\rm V}_{\rm bias}^{(j)}(Q)\right),
\label{eq:NIDef}
\end{eqnarray}
assuming that both ${\rm P}(Q)$ and ${\rm P}^{(j)}_\text{bias}(Q)$ are normalized. However, this estimate for ${\rm P}(Q)$ will only be useful for $Q$ bins that have good statistics in the $j$-th simulation, which in our case means that the bins are at an $r$ coordinate that is close to the $r_j$ value of the respective bias. Another problem with this expression is that in order to calculate $N^{(j)}$ we already need to know the sought-for quantity ${\rm P}(Q)$. To deal with the first problem, we combine the individual estimates obtained from each $j$-th simulation, to form an improved estimate:
\begin{eqnarray}
{\rm P}_{\rm est}(Q)=\sum_j c^{(j)}(Q) {\rm P}_{{\rm est},j}(Q).
\label{eq:pEst0Def}
\end{eqnarray}
With $c^{(j)}(Q)$ we can tune the weight of ${\rm P}_{{\rm est},j}(Q)$ in the ${\rm P}(Q)$ estimate. We require $\sum_j c^{(j)}(Q)=1$. In bins where the $j$-th simulation has bad statistics we will choose $c^{(j)}(Q)$ close to $0$, such that the ${\rm P}_{{\rm est},j}(Q)$ estimate contributes only in bins where it is useful. The error of ${\rm P}_{\rm est}(Q)$ can be estimated via the Poisson distribution and it can be minimized via the following choice for the $c^{(j)}(Q)$:
\begin{eqnarray}
c^{(j)}(Q)= \frac{\exp\left(-\beta {\rm V}_{\rm bias}^{(j)}(Q) \right) M^{(j)} N^{(j)}}{\sum_k \exp\left(-\beta {\rm V}_{\rm bias}^{(k)}(Q) \right) M^{(k)} N^{(k)}}.
\label{eq:cIDef}
\end{eqnarray}
Here $M^{(j)}$ is the number of uncorrelated configurations sampled in the $j$-th simulation. With Eqs. \ref{eq:pEst0IDef}-\ref{eq:cIDef} we now arrive at an expression for an estimate of ${\rm P}(Q)$. However, as $N^{(j)}$ depends on ${\rm P}(Q)$ the expression (\ref{eq:NIDef}) can only be evaluated if ${\rm P}(Q)$ is known in the first place. We can deal with this problem by using ${\rm P}_{\rm est}(Q)$ for ${\rm P}(Q)$ in the formula for $N^{(j)}$ (\ref{eq:NIDef}) to obtain a self-consistency problem for ${\rm P}_{\rm est}(Q)$, which can be solved iteratively. As a starting point for this iterative procedure an initial guess for ${\rm P}_{\rm est}(Q)$ has to be provided. However, after many iterations the procedure is expected to converge to the same distribution, independent of the given initial condition. We started the iterative algorithm with an uniform distribution for ${\rm P}_{\rm est}(Q)$.

\section{Expansion of ${\rm g}\left(r,\cos\theta_1,\cos\theta_2,\varphi\right)$}
\label{App:expansion}
As the anisotropic effective potential ${\rm V_{eff}}\left(r,\cos\theta_1,\cos\theta_2,\varphi\right)$ and the corresponding pair-correlation function ${\rm g}\left(r,\cos\theta_1,\cos\theta_2,\varphi\right)$ depend on 4 variables, it is hard to visualize them. Nevertheless we can obtain a quantitative measure of the anisotropy in ${\rm g}$ and ${\rm V_{eff}}$ by carrying out an expansion of g:
\begin{eqnarray}
{\rm g}\left(r,\cos\theta_1,\cos\theta_2,\varphi\right)= \sum_{n} {\rm c}_{n}(r) \,{\rm f}_{n}\left(\cos\theta_1,\cos\theta_2,\varphi\right).
\end{eqnarray}
Here, ${\rm f}_{n}$ are modes that depend on the angular degrees of freedom only and they form a complete basis for the angular dependence of $\rm g$.

To obtain a suitable set of basis functions ${\rm f}_n$ for the dependence of $\rm g$ on the directors $\mathbf{d}^{(1,2)}$ for a given vector $\mathbf{r}$ between the rings, we started with an expansion to a sum of products of spherical harmonics:
\begin{eqnarray}
\gamma\left(\mathbf{d}^{(1)},\mathbf{d}^{(2)}\right)= \sum_{l_1=0}^\infty \sum_{m_1=-l_1}^{l_1} \sum_{l_2=0}^\infty \sum_{m_2=-l_2}^{l_2} c_{l_1,m_1,l_2,m_2} Y_{l_1}^{m_1}\left(\cos\theta_1,\varphi_1\right) Y_{l_2}^{m_2}\left(\cos\theta_2,\varphi_2\right).
\end{eqnarray}
Here $\gamma$ denotes $\rm g$ at a given $\mathbf{r}$. By $\cos\theta_i$ and $\varphi_i$ the director $\mathbf{d}^{(i)}$ is represented in spherical coordinates. We use a reference frame where the connection vector $\mathbf{r}$ between the two rings points to the north-pole and therefore $\varphi_i$ denotes the azimuthal angle around $\mathbf{r}$. With $Y_{l}^{m}$ we denote the spherical harmonics\cite{axler2001harmonic}. They fulfil the orthonormality relation $\int d \Omega\, \bar{Y}_{l}^m(\cos\theta,\varphi)\,Y_{l'}^{m'}(\cos\theta,\varphi)= \delta_{l,l'} \delta_{m,m'}$ with $d\Omega=d\cos\theta d\varphi$, where $\bar{z}$ denotes the complex conjugation of a complex number $z$. This allows us to compute the expansion coefficients via integration:
\begin{eqnarray}
\label{eq:coeffByIntOld}
c_{l_1,m_1,l_2,m_2}=\int d \Omega_1 d \Omega_2 \,\bar{Y}_{l_1}^{m_1}(\cos\theta_1,\varphi_1)\,\bar{Y}_{l_2}^{m_2}(\cos\theta_2,\varphi_2)\,\gamma(\cos\theta_1,\varphi_1,\cos\theta_2,\varphi_2).
\end{eqnarray}

Due to the symmetries of $\gamma$, we will now be able to compute or relate many of the coefficients and thus arrive at a reduced set of basis functions $\lbrace{\rm f}_{l_1,l_2,m} \rbrace$ with which we can still represent $\gamma$ exactly. We first use the continuous symmetry under rotations around the connection vector $\mathbf{r}$:
\begin{eqnarray}
\label{eq:azimuthRot}
\gamma(\cos\theta_1,\varphi_1,\cos\theta_2,\varphi_2)=\gamma(\cos\theta_1,\varphi_1-\varphi_2,\cos\theta_2,0).
\end{eqnarray}
Using (\ref{eq:coeffByIntOld}) and $Y_{l}^m(\cos\theta,\varphi)\propto \exp(i m \varphi)$ one can show that $c_{l_1,m_1,l_2,m_2}$ vanishes for $m_1\neq-m_2$ due to this symmetry. In the following, we enumerate the expansion modes with $m\equiv m_1=-m_2$. Next we use that $\mathbf{d}^{(i)}$ is equivalent to $-\mathbf{d}^{(i)}$ and therefore $\gamma$ fulfils the symmetry $\gamma\left(\mathbf{d}^{(1)},\mathbf{d}^{(2)}\right)=\gamma\left(-\mathbf{d}^{(1)},\mathbf{d}^{(2)}\right)=\gamma\left(\mathbf{d}^{(1)},-\mathbf{d}^{(2)}\right)$. Since $Y_{l}^m(\mathbf{d})= (-1)^l Y_{l}^m(-\mathbf{d})$, $c_{l_1,m_1,l_2,m_2}$ are zero if either $l_1$ or $l_2$ are odd. 

The monomer-resolved model is symmetric under a mirror transformation, which is therefore also a symmetry of the effective model. If we consider a state with $\varphi_2=0$ and mirror it by a plane spanned by $\mathbf{r}$ and $\mathbf{d}^{(2)}$ we obtain a state with identical $\mathbf{r}$, $\mathbf{d}^{(2)}$ and $\cos\theta_1$, while $\varphi_1$ changes sign. Hence we obtain:
\begin{eqnarray}
\label{eq:phi12Swap}
&&\gamma(\cos\theta_1,\varphi_1,\cos\theta_2,\varphi_2)=\gamma(\cos\theta_1,\varphi_1-\varphi_2,\cos\theta_2,0)=\nonumber\\
&&\gamma(\cos\theta_1,\varphi_2-\varphi_1,\cos\theta_2,0)=\gamma(\cos\theta_1,\varphi_2,\cos\theta_2,\varphi_1).
\end{eqnarray}
For the first and the last step we used (\ref{eq:azimuthRot}). Therefore $\gamma$ is invariant under exchanging $\varphi_1$ and $\varphi_2$. We therefore have  $c_{l_1,m,l_2,-m}=c_{l_1,-m,l_2,m}$.
%Therefore $\gamma$ is even in $\varphi_1-\varphi_2$. We can span the two dimensional space of $Y_{l_1}^{m}Y_{l_2}^{-m} \sim \exp(i m (\varphi_1-\varphi_2))$ and  $Y_{l_1}^{-m}Y_{l_2}^{m} \sim \exp(i m (\varphi_2-\varphi_1))$ with the two functions $f_1\equiv Y_{l_1}^{m}Y_{l_2}^{-m} +Y_{l_1}^{-m}Y_{l_2}^{m}\sim \cos(\varphi_1-\varphi_2)$ and $f_2\equiv Y_{l_1}^{m}Y_{l_2}^{-m} - Y_{l_1}^{-m}Y_{l_2}^{m}\sim \sin(\varphi_1-\varphi_2)$. Since $f_2$ is uneven in $\varphi_1-\varphi_2$ its coefficient in an expansion of $\gamma$ will vanish. Therefore we have $c_{l_1,m,l_2,-m}=c_{l_1,-m,l_2,m}$

Everything discussed so far also holds if we calculate the effective interaction between different rings, e.g. for rings with a different number of monomers. The final symmetry, which we will exploit now, only holds if we have identical rings. In this case we  obtain an equivalent state if we swap the orientations of the two ring polymers:
\begin{eqnarray}
\gamma(\cos\theta_1,\varphi_1,\cos\theta_2,\varphi_2)= \gamma(\cos\theta_2,\varphi_2,\cos\theta_1,\varphi_1)=\gamma(\cos\theta_2,\varphi_1,\cos\theta_1,\varphi_2)
\end{eqnarray}
For the last transformation we used (\ref{eq:phi12Swap}). Hence for identical rings $\gamma$ is also invariant under an exchange of $\theta_1$ and $\theta_2$ and we therefore know that  $c_{l_1,m,l_2,-m}=c_{l_2,m,l_1,-m}$.

We now group basis functions if we a priori know that $\gamma$ has identical coefficients $c_{l_1,m,l_2,-m}$ with respect to them. We sum the modes in each group and divide by $\sqrt{\#}$, where $\#$ is the number of modes in the group. For identical rings $\#$ is at most 4, but can also be smaller if $l_1=l_2$ or $m=0$. For different rings $\#= 2$ if $m\neq0$ and 1 otherwise. In this way we obtain a new set of basis functions $f_{l_1,l_2,m}$. The indices of $f_{l_1,l_2,m}$ refer to the indices of one of the modes in the group from which $f_{l_1,l_2,m}$  was constructed, with the additional constraint that $m\geq 0$ and $l_2\geq l_1$ in the case of identical rings. As an example we can consider $f_{2,4,1}$, which is set to $\left(Y_{2}^{1} Y_{4}^{-1}+Y_{2}^{-1} Y_{4}^{1}+Y_{4}^{1} Y_{2}^{-1}+Y_{4}^{-1} Y_{2}^{1}\right)/\sqrt{4}$ in the case of identical and $\left(Y_{2}^{1} Y_{4}^{-1}+Y_{2}^{-1} Y_{4}^{1}\right)/\sqrt{2}$ in the case of different rings. The new basis functions $f_{l_1,l_2,m}$ only depend on $\varphi\equiv |\varphi_1-\varphi_2|$ and not on $\varphi_1$, $\varphi_2$ separately. Thus $|\varphi_1-\varphi_2|$ is precisely the $\varphi$ coordinate defined in (\ref{eq:effCoordsDef}), on which g depends. Expanding the angular dependence of g with our new basis functions, we obtain
\begin{eqnarray}
\label{eq:expansionFormula}
{\rm g}\left(r,\cos\theta_1,\cos\theta_2,\varphi\right)= \sum_{l_1=\lbrace 0,2,4,...\rbrace} \sum_{l_2= \lbrace l_1,l_1+2,l_1+4,... \rbrace} \sum_{m=0}^{\text{min}(l_1,l_2)} {\rm c}_{l_1,l_2,m}(r) \,{\rm f}_{l_1,l_2,m}\left(\cos\theta_1,\cos\theta_2,\varphi\right).
\end{eqnarray}
In the case of different rings, the sum over $l_2$ does not start at $l_1$ but at 0. Like $\lbrace Y_{l_1}^{m} Y_{l_2}^{-m} \rbrace$, also $\lbrace f_{l_1,l_2,m}\rbrace$ fulfil the orthonormality relation
\begin{eqnarray}
\int d \Omega_1 d \Omega_2 \, f_{l_1,l_2,m}(\cos \theta_1, \cos \theta_2, \varphi) f_{l'_1,l'_2,m'}(\cos \theta_1, \cos \theta_2, \varphi)= \delta_{l,l'} \delta_{m,m'}.
\end{eqnarray}
Note that complex conjugation is not necessary as $f_{l_1,l_2,m}$ are real functions in contrast to $Y_{l_1}^{m} Y_{l_2}^{-m}$. Hence, we can obtain the coefficients ${\rm c}_{l_1,l_2,m}(r)$ via an integration analogous to (\ref{eq:coeffByIntOld}):
\begin{eqnarray}
\label{eq:coeffByIntNew}
c_{l_1,l_2,m}(r)=\int d \Omega_1 d \Omega_2 \,f_{l_1,l_2,m}(\cos \theta_1, \cos \theta_2, \varphi)\,{\rm g} (r,\cos\theta_1,\cos\theta_2,\varphi).
\end{eqnarray}
To calculate $c_{l_1,l_2,m}(r)$ we could do a numerical integration of ${\rm g}(r,\cos\theta_1,\cos\theta_2,\varphi)$.
%$\rm g$ is related to ${\rm P}(r,\cos \theta_1, \cos_2, \varphi)$ via equation (\ref{eq:gPRelation}). As described in section \ref{sec:SelfConsistentHisto} we calculate $\rm P$ numerically on a 4D grid using \textit{Umbrella-Sampling} and the \textit{Self-Consistent Histogram Method}.
A different approach for the calculation of $c_{l_1,l_2,m}(r)$ is to express the integral in (\ref{eq:coeffByIntNew}) as an average over configurations of a system of 2 ring polymers for fixed $r$. We sample these configurations from the simulations with the bias potential ${\rm V_{bias}}(r)$ given in (\ref{eq:VBias}), which we carried out for calculating $\rm g$ on a 4D grid using \textit{umbrella sampling}. As ${\rm V_{bias}}(r)$ does not change the relative weight of configurations with identical $r$ values, we can use the configurations that fall in a small window of $r$ values to estimate averages over the angular degrees of freedom at some fixed $r$ value. The average of the expansion modes $f_{l_1,l_2,m}$ over the configurations of the rings is related to $c_{l_1,l_2,m}(r)$ as follows:
\begin{eqnarray}
\langle f_{l_1,l_2,m}\rangle_r&=& \frac{\int d\Omega_1 d\Omega_2 \, {\rm g}(r,\cos\theta_1,\cos\theta_2,\varphi) \, f_{l_1,l_2,m}(\cos \theta_1, \cos \theta_2, \varphi)}{\int d\Omega_1 d\Omega_2 \, {\rm g}(r,\cos\theta_1,\cos\theta_2,\varphi)}=\nonumber\\
&=&\frac{c_{l_1,l_2,m}(r)}{(4 \pi)^2 {\rm g^{iso}}(r)}.
\label{eq:cByAverage}
\end{eqnarray}
Here we used (\ref{eq:gIsoGRelation}) and (\ref{eq:coeffByIntNew}) for the final step. $\rm g^{iso}$ is easy to calculate numerically once we know $\rm g$. The advantage of this approach with respect to a numerical evaluation of (\ref{eq:coeffByIntNew}) is that we do not need to introduce a grid in the angular degrees of freedom. In particular if we want to calculate $c_{l_1,l_2,m}(r)$ coefficients for high $l_1$, $l_2$ values the correctness of the result in the first approach depends sensitively on the number of grid points and also on the numerical method for carrying out the integration. Calculating averages on the other hand is straight-forward and we can easily estimate the statistical error using the standard deviation of block-averages. For these reasons we calculate the expansion coefficients $c_{l_1,l_2,m}(r)$ and estimate their errors following the latter approach.

The larger $l_i$, the faster the $f_{l_1,l_2,m}$ functions can change when the director $\mathbf{d}^{(i)}$ is varied. $f_{0,0,0}=(4\pi)^{-1}$ is constant and therefore gives the isotropic contribution. By comparing ${\rm c}_{0,0,0}(r)$ with the size of the coefficients for $(l_1,l_2,m)\neq (0,0,0)$ we can quantify the importance of anisotropy in the effective interaction.

We find that with the 14 coefficients ${\rm c}_{l_1,l_2,m}(r)$ for which $l_1,l_2\leq 4$ we already get a reasonable approximation of $\rm g$. This is true, even for the smallest rings investigated ($N=20$), where the anisotropy is most important. This fact allows us to store the essential information in the anisotropic potential with only a few functions of one variable, $c_{l_1,l_2,m}(r)$, instead of a 4D grid with a very large number of grid points. Using only the coefficient ${\rm c}_{0,0,0}(r)$ we recover ${\rm g^{iso}}(r)$ of the isotropic effective model.

\bibliography{bibliography}

\end{document}